\documentclass[onecolumn,sort&compress,numbers]{els-mrw} % For numbered references Style 

\usepackage{amsmath,amssymb,amsfonts,amsthm,makeidx,graphicx}
\usepackage{txfonts}
\usepackage{helvet}
\usepackage{color}
\usepackage{slashed}
\usepackage{hyperref}
\usepackage{bm}
\usepackage{multirow}
\usepackage{array,booktabs,colortbl,multirow}
\usepackage{xcolor,colordvi}
\usepackage{rotating}
\usepackage{placeins}
\begin{document}

%%%%%%%%%%%%%%%%%%%%%%%%%%%%%%%%%%%%%%%%%%%%%%%%%%%%%%%%%%%%%%%%
%% the following items are mandatory: 
%% - title
%% - author names
%% - affiliation details
%% - abstract
%% - keywords

%% Precise, concise, and informative description of the focus of this work. Avoid abbreviations and formulae in the title
\chapter{Electroweak precision tests}
\label{chap1}

%% All author names and affiliations, and email address for corresponding author
\author[1]{Laura Reina}%
\author[2]{Luca Silvestrini}%

\address[1]{\orgname{Florida State University}, \orgdiv{Physics
    Department}, \orgaddress{77 Chieftan Way, Tallahassee, FL
    32306-4350, USA}}
\address[2]{\orgname{INFN}, \orgdiv{Sezione di Roma}, \orgaddress{Piazzale A. Moro 2, I-00185 Rome, Italy}}

%\articletag{Chapter Article tagline: update of previous edition, reprint.}

\maketitle

%%%%%%%%%%%%%%%%%%%%%%%%%%%%%%%%%%%%%%%%%%%%%%%%%%%%%%%%%%%%%%%%
%% the following item is mandatory: 
%% 100-150 word summary of the chapter
\begin{abstract}[Abstract]
  %	We give a pedagogical introduction...(100-150 words)
 The Standard Model of particle physics provides a rigorous framework
 within which processes mediated by electroweak interactions can be
 calculated with great accuracy. By comparing with high-precision
 experimental measurements of the same processes, deviations from
 Standard Model predictions can be identified as indirect signals of
new physics. In particular, electroweak precision fits combine multiple observables
and provide a unique test of the Standard Model consistency at the quantum level.
\end{abstract}

%% 5-10 words that embody the key topics in the chapter. What terms would someone put into a search engine if they were looking for a chapter like this?
\begin{keywords}
% 	please enter 5 keywords as follows:
 	Standard Model \sep electroweak gauge bosons \sep Higgs
        boson\sep precision electroweak observables \sep precision electroweak fits 
\end{keywords}

%%%%%%%%%%%%%%%%%%%%%%%%%%%%%%%%%%%%%%%%%%%%%%%%%%%%%%%%%%%%%%%%
%% the following item is optonal: 
%% - Single figure visually illustrating the key topic/method/outcome described in the chapter
%\begin{figure}[h]
%	\centering
%	\includegraphics[width=7cm,height=4cm]{blankfig}
%	\caption{Optional: Single figure visually illustrating the key topic/method/outcome described in the chapter. 
%		     Please add here some text explaining the pic...}
%	\label{fig:titlepage}
%\end{figure}

%%%%%%%%%%%%%%%%%%%%%%%%%%%%%%%%%%%%%%%%%%%%%%%%%%%%%%%%%%%%%%%%
%% the following item is optional: 
%% - System of abbreviations/terms/symbols used in the specific field of study/community. List and define
\begin{glossary}[Nomenclature]
	\begin{tabular}{@{}lp{34pc}@{}}
          SM & Standard Model\\
          EW & Electroweak\\
          SSB & Spontaneous Symmetry Breaking\\
          EWPO & Electroweak Precision Observables\\
          LHC & Large Hadron Collider\\
          NP & New Physics\\
          SMEFT & Standard Model Effective Field Theory
	\end{tabular}
\end{glossary}

%%%%%%%%%%%%%%%%%%%%%%%%%%%%%%%%%%%%%%%%%%%%%%%%%%%%%%%%%%%%%%%%
%% the following item is mandatory: 
%% List of the key points and topics a reader can expect to learn from this chapter 
\section*{Objectives}
\begin{itemize}
	\item Concise review of the Standard Model electroweak sector.
	\item Introduction to electroweak precision observables, theoretical and experimental state of the art.
	\item Updated electroweak precision fit in the Standard Model.
	\item Examples of constraints on new physics from electroweak precision tests and global fits of collider measurements.
\end{itemize}

%%%%%%%%%%%%%%%%%%%%%%%%%%%%%%%%%%%%%%%%%%%%%%%%%%%%%%%%%%%%%%%%
%% the following items are mandatory: 
%% - Section: Introduction 
%% - further sections
%% - Section: Conclusion
\section{Introduction}
\label{ewpf:sec:intro}
%Please provide a very general and easy to understand introduction to your chapter.

The nature and properties of electroweak interactions are encoded in
the Standard Model (SM) of particle physics, the quantum field theory
that rigorously describes all elementary-particle interactions, with
the exception of gravity whose effects are negligible at subatomic
scales. The
scientific exploration that led to formulate the portion of the SM
that describes electroweak (EW) interactions is one of the richest chapters
in the history of particle physics. The emerging evidence of a new
kind of \textit{weak} force in studies of radioactivity and
$\beta$-decay in the early $20^{\mathrm{th}}$ century led in the
1930s to the formulation of the famous Fermi theory of weak
interactions~\cite{fermi:1934zfp}, a four-fermion interaction involving a contact
force that would be later on explained as due to the exchange of a
very massive mediator (the $W$ boson).  Several revolutionary
developments followed in the span of just a few decades.  In the
1950s, the suggestion~\cite{Lee:1956qn} and subsequent discovery~\cite{Wu:1957my} of
parity violation in weak interactions defined the chiral nature of
weak interactions, while the generalization of gauge theories from
abelian to non-abelian symmetries~\cite{Yang:1954ek} introduced the
possibility of describing all interactions besides electromagnetic interactions as
mediated by the exchange of vector (spin-1) particles like the photon of electromagnetic
interactions. What would be later on labeled as \textit{Standard Model}
emerged in the 1960s~\cite{Glashow:1961tr,Weinberg:1967tq,Salam:1968rm} and proposed the unification of weak and
electromagnetic interactions in terms of a spontaneously broken gauge
theory based on the local $\mathrm{SU}(2)_L\times \mathrm{U}(1)_Y$
chiral symmetry group  (associated to quantum numbers of weak isospin, $I_3$,
and hypercharge, $Y$, respectively) spontaneously broken to only preserve the gauge
symmetry of quantum electrodynamics (QED), namely $\mathrm{U}(1)_{QED}$~\cite{Englert:1964et,Higgs:1964pj}. Besides
the photon as mediator of electromagnetic interactions, the SM
predicted the existence of massive charged ($W^\pm$) and neutral ($Z$)
mediators of weak interactions, as well as  a neutral scalar (spin-0) particle,
the Higgs boson ($H$), with a specific pattern of masses and
couplings. The universality of weak interactions, apparently violated in the hadronic sector, was restored with the introduction of the Cabibbo angle \cite{Cabibbo:1963yz}. The absence of tree-level flavour-changing neutral currents, and their loop-level suppression, was then explained via the GIM mechanism \cite{Glashow:1970gm}, allowing for two \textit{families} of quarks to fit in the SM together with leptons. The observation of CP violation in Kaon decays \cite{Christenson:1964fg} suggested the presence of a third family of quarks to allow for a CP-violating phase in the Cabibbo-Kobayashi-Maskawa mixing matrix \cite{Kobayashi:1973fv}. With the discovery of the $W^\pm$~\cite{UA1:1983crd,UA2:1983tsx} and
$Z$ bosons~\cite{UA1:1983mne,UA2:1983mlz}, the confirmation of the existence
of three complete families of fermions after the discovery of the top
quark~\cite{CDF:1995wbb,D0:1995jca},
and the most recent discovery of the Higgs boson~\cite{Aad:2012tfa,Chatrchyan:2012xdj}, we have strong
evidence that the SM is the theory describing all known elementary
particles and their interactions at what is generically identified as
\textit{electroweak} scale, an energy scale of a few hundreds GeV 
defined by the mass scale of EW interactions.

Most importantly, the SM EW theory has been shown to be a full fledged
renormalizable perturbative quantum field theory~\cite{tHooft:1972tcz}
capable of providing unambiguous physical predictions that only depend
on a small finite number of independent input parameters (masses and
couplings) and can be systematically improved as more and more terms
are calculated in the theory perturbative expansion.\footnote{The
  perturbative parameters of the SM quantum field theory are the
  couplings of the electroweak and strong interactions that define the
  theory.} In particular, the SM allows to obtain highly accurate
predictions for a set of physical observables mainly related to the
properties of the $Z$ and $W$ bosons for which very precise
measurements have been obtained at both lepton ($e^+e^-$) and hadron
($p\bar{p},pp$) colliders over the last several
decades~\cite{ParticleDataGroup:2024cfk}, from the LEP (CERN) and SLC
(SLAC) $e^+e^-$ colliders~\cite{ALEPH:2005ab,LEP-2}, to the Tevatron
(FNAL) $p\bar{p}$
collider~\cite{ALEPH:2010aa,TevatronElectroweakWorkingGroup:2010mao,D0:2012kms,
  CDF:2016cry,CDF:2016vzt,CDF:2022hxs}, and more recently the $pp$
Large Hadron Collider (LHC,
CERN)~\cite{LHCb:2021bjt,ATLAS:2024erm,CMS:2024lrd,ATLAS:2022jbw,ATLAS:2018fwq,CMS:2023ebf,CMS:2022kqg,ATLAS:2023oaq,CMS:2024eka, Aaij:2015lka,LHCb:2024ygc,Aad:2015uau,ATLAS:2018gqq,Sirunyan:2018swq,CMS:2024ony,CMS:2023mgq,CMS:2022mhs}.
They are the so-called \textit{Electroweak Precision Observables}
(EWPO) which comprise observables such as $Z$- and $W$-boson masses and
widths, $Z$-boson decay rates at the $Z$ resonance, rates of $Z$- and
$W$-boson induced processes, and various other observables defined in
terms of these fundamental building blocks as well as quantities that
are not specific of the EW sector but enter the predictions of EWPO.
%Electroweak Precision Observables (EWPO) consist of a set of
%quantities related to EW gauge-bosons masses and couplings, that can
%be calculated with extreme accuracy and for which very precise
%measurements have been obtained at both lepton ($e^+e^-$) and hadron
%($p\bar{p},pp$) colliders over the last several
%decades~\cite{ParticleDataGroup:2024cfk}, from the LEP (CERN) and SLC
%(SLAC) $e^+e^-$ colliders~\cite{ALEPH:2005ab}, to the Tevatron (FNAL)
%$p\bar{p}$ collider~\cite{ALEPH:2010aa,CDF:2016cry,CDF:2016vzt}, and
%more recently the $pp$ Large Hadron Collider (LHC, CERN)~\cite{Khachatryan:2015hba,Aaboud:2018zbu,Sirunyan:2018gqx,Sirunyan:2018goh,Sirunyan:2018mlv,ATLAS:2019ezb,Aaboud:2017svj,Aad:2015uau,ATLAS:2018gqq,Sirunyan:2018swq,Aaij:2015lka}.
%They comprise obesrvables such as $Z$ and $W$ boson masses and
%widths, $Z$-boson decay rates at the $Z$ resonance, rates of $Z$ and
%$W$-boson induced processes, and various other observables defined in terms of these fundamental building blocks.

EW precision tests compare SM predictions for EWPO to their
measurements and, given the correlations induced by the dependence of
such predictions on a common set of input parameters, they can probe
the consistency of the SM theory. Since both EWPO theoretical
predictions and experimental measurements have reached high accuracy, EW
precision tests have a unique capability to \text{stress-test} the
SM and either point to its missing components or to new physics (NP). They have a long
history~\cite{Kim:1980sa,Amaldi:1987fu,Kennedy:1990ib} and in the past
thirty years have played a key role in predicting the mass regions
where the last missing building blocks of the SM, the top-quark and
the Higgs boson, were eventually
discovered~\cite{Erler:1994fz,ParticleDataGroup:1994kdp,Flacher:2008zq,Baak:2011ze,LEPEWWG}. Having
discovered all the elementary constituents of the SM and having
measured their properties with high precision, EW precision tests of
the SM can now be used to reveal possible tensions that may be
resolved in models of physics beyond the SM. Indeed EW precision tests
represent to these days one of the strongest constraints for models of
NP~\cite{Costa:1987qp,Langacker:1991an,Peskin:1991sw,Altarelli:1993sz,Altarelli:1991fk,Grinstein:1991cd,Barbieri:1999tm,Barbieri:2004qk,Erler:2013xha,Ciuchini:2013pca,Akhundov:2013ons,Baak:2014ora,deBlas:2016ojx,Haller:2018nnx,Freitas:2020kcn,deBlas:2021wap,deBlas:2022hdk}.\footnote{For recent comprehensive reviews of EW precision tests see also~\cite{Erler:2019hds,Freitas:2020kcn}.}

In this Chapter we will review the framework and key
ingredients of EW precision tests and present state-of-the-art results for EW precision fits of the SM and general models of NP. After having provided a
brief introduction to the SM with emphasis on the EW sector
in Section~\ref{ewptests:sec:sm-lagrangian}, we will introduce the set of
EWPO considered in EW precision fits and review the
state-of-the-art of both their theoretical predictions and
experimental measurements in
Section~\ref{ewptests:sec:ew-precision-observables}. We will then discuss more technical details of EW precision fits in Section~\ref{ewptests:sec:ew-precision-fits}
and present a comprehensive set of results of EW precision fits of the SM, including all most recent theoretical and experimental results, that will illustrate the
current degree of consistency of the SM. Finally, in Section~\ref{ewptests:sec:ew-precision-fits-bsm} we
will discuss the constraints imposed by EW precision fits on
theories beyond the SM and highlight how the precision currently reached by the LHC experiments in
the measurement of a much broader set of observables, including
top-quark and Higgs-boson observables, has allowed to extend the idea
of EW precision fits to more general global fits that, given
the precision attainable in the high-luminosity phase of the LHC
(HL-LHC) as well as at future colliders, have the potential to
provide indirect evidence of NP and guide future discoveries.

%%%%%%%%%%%%%%%%%%%%%%%%%%%%%%%%%%%%%%%%%%%%%%%%%%%%%%%%%%%%%%%%
%% in the following we showcase possible style elements
%%
%%
\section{The Standard Model Lagrangian}
\label{ewptests:sec:sm-lagrangian}
The SM is a quantum field theory invariant under the $\mathrm{SU}(3)_C\otimes \mathrm{SU}(2)_L\otimes \mathrm{U}(1)_Y$ local or \textit{gauge} symmetry group. The $\mathrm{SU}(3)_C$ gauge symmetry defines Quantum Chromodynamics (QCD), the theory governing the dynamics of strong interactions while the $\mathrm{SU}(2)_L\otimes \mathrm{U}(1)_Y$ gauge symmetry determines the dynamics of EW interactions. The Lagrangian of the SM  can be written in a compact and manifestly gauge-invariant form as:
\begin{equation}
\begin{split}
    \mathcal{L}_\mathrm{SM} = &-\frac{1}{4}G_{\mu\nu}^A G^{A,\mu\nu} 
                             -\frac{1}{4}W_{\mu\nu}^I W^{I,\,\mu\nu} 
                             -\frac{1}{4}B_{\mu\nu} B^{\mu\nu}
                             + (D_{\mu}\phi)^{\dagger}(D^{\mu}\phi) 
                             - \mu^2\phi^{\dagger}\phi
                             - \lambda(\phi^{\dagger}\phi)^2 \\
                             &+ i\left( \bar{l}^i_L \slashed{D} l^i_L + \bar{e}^i_R \slashed{D} e^i_R 
                                       + \bar{q}^i_L \slashed{D} q^i_L + \bar{u}^i_R \slashed{D} u^i_R +\bar{d}^i_R \slashed{D} d^i_R\right) \\
                             &- \left( \bar{l}^i_L \Gamma_e e^i_R \phi + \bar{q}^i_L \Gamma_u u^i_R \tilde\phi 
                                      + \bar{q}^i_L \Gamma_d d^i_R \phi  \right) + h.c.\,,
\end{split}
\label{ewptests:eq:sm-lagrangian}
\end{equation}
where the first line of Eq.~(\ref{ewptests:eq:sm-lagrangian})
encapsulates the dynamics of the gauge fields and their interactions
with the scalar Higgs field ($\phi$), while the second line defines the dynamics
of all fermion fields (leptons and quarks) and their interactions with
the gauge fields (summed over three fermion generations for $i=1,2,3$). Left-handed lepton and quark $\mathrm{SU}(2)_L$ doublets have been denoted by $l^i_L$ and $q^i_L$, while right-handed $\mathrm{SU}(2)_L$ singlets by $e^i_R$, $u^i_R$, and $d^i_R$ respectively. The
last line introduces Yukawa-type interactions between the fermion
fields and the scalar Higgs field, with $\tilde\phi=i\sigma^2\phi^*$. 
The covariant derivative used in Eq.~(\ref{ewptests:eq:sm-lagrangian}) is defined as:
\begin{equation}
    D_{\mu}=\partial_{\mu}+ig_sG_{\mu}^A\mathcal{T}^A+ig_2 W_{\mu}^I T^I+ig_1B_{\mu}Y\,,
\label{ewptests:eq:CovariantDerivative}
\end{equation}
in terms of the gauge fields $G_{\mu}^A$, $W_{\mu}^I$, and $B_{\mu}$  associated with the strong ($\mathrm{SU}(3)_C$), weak isospin ($\mathrm{SU}(2)_L$), and hypercharge ($\mathrm{U}(1)_Y$) interactions respectively, with coupling constants $g_s$, $g_1$, and $g_2$, while $\textit{T}^I = \tau^I/2$ are the generators of the $\mathrm{SU}(2)$ group expressed in terms of Pauli matrices $\tau^I$ (I=1,2,3),  and $\mathcal{T}^A=\lambda^A/2$ are the generators of the $\mathrm{SU}(3)$ group expressed in terms of Gell-Mann matrices $\lambda^A$ (A=1,...,8). The corresponding field strength tensors are defined as:
\begin{subequations}
\begin{align}
    G_{\mu\nu}^A &= \partial_{\mu}G_{\nu}^A - \partial_{\nu}G_{\mu}^A - g_s f^{ABC}G_{\mu}^B G_{\nu}^C\,, \\
    W_{\mu\nu}^I &= \partial_{\mu}W_{\nu}^I - \partial_{\nu}W_{\mu}^I - g_2 \epsilon^{IJK}W_{\mu}^J W_{\nu}^K\,, \\
    B_{\mu\nu} &= \partial_{\mu}B_{\nu} - \partial_{\nu}B_{\mu} \,,
\end{align}
\end{subequations} 
where $f^{ABC}$ and $\epsilon^{IJK}$ are the structure constants of
the non-abelian $\mathrm{SU}(3)$ and $\mathrm{SU}(2)$ groups. The complex scalar field
$\phi$ is the $\mathrm{SU}(2)_L$ doublet (singlet of $\mathrm{SU}(3)_C$) that is responsible for the spontaneous symmetry
breaking (SSB) of the EW theory through the Brout-Englert-Higgs mechanism
\cite{Englert:1964et,Higgs:1964pj}. The
fermion fields from the Dirac and Yukawa part of the Lagrangian
describe the three generations of quarks and leptons with quantum
numbers as summarized in Table
\ref{ewptests:tab:fermion-quantum-numbers}.

\begin{table}[t]
\centering
\small
\renewcommand{\arraystretch}{1.5}
\addtolength{\tabcolsep}{-0.1em}
    \begin{tabular}{||c||c|c|c|c|c|c|c||}
        \hline \hline
             & $\nu_L^i$ & $e_L^i$ & $e_R^i$ & $u_L^i$ & $d_L^i$ & $u_R^i$ & $d_R^i$  \\
         \hline \hline 
         C  &  0 & 0 & 0 & 3 & 3 & 3 & 3 \\
         $I_3$ & $\frac{1}{2}$ & -$\frac{1}{2}$ & 0 & $\frac{1}{2}$ & -$\frac{1}{2}$ & 0 & 0\\
         Y   & -$\frac{1}{2}$ & -$\frac{1}{2}$ & -1 & $\frac{1}{6}$ & $\frac{1}{6}$ & $\frac{2}{3}$ & -$\frac{1}{3}$ \\
         Q   & 0 & -1 & -1 & $\frac{2}{3}$ & -$\frac{1}{3}$ & $\frac{2}{3}$ & -$\frac{1}{3}$ \\
         \hline \hline
    \end{tabular}
    \caption{Color (C), weak isospin ($I_3$), hypercharge (Y), and charge (Q) quantum numbers of the left and right-handed fermions of the $i^{\mathrm{th}}$ family ($i=1,2,3$).}
\label{ewptests:tab:fermion-quantum-numbers}
\end{table}

In this Section we focus on the EW part of the SM
Lagrangian, and review those aspects of the EW SM that are most
important to define EWPO and discuss EW precision tests.
For more comprehensive introductions to QCD and the full SM
we refer to the corresponding dedicated chapters in this Encyclopedia.
The EW gauge symmetry is realized as a spontaneously broken
symmetry. While the EW Lagrangian in
Eq.~\ref{ewptests:eq:sm-lagrangian} is invariant under
$\mathrm{SU}(2)_L\times \mathrm{U}(1)_Y$, the ground (or vacuum) state of the theory is
not. Indeed, for $\mu^2<0$, the scalar potential of
Eq.~\ref{ewptests:eq:sm-lagrangian},
$V(\phi)=\mu^2\phi^\dagger \phi+\lambda(\phi^\dagger\phi)^2$, develops
a non-zero global minimum at
$\langle\phi^\dagger\phi\rangle_0=\varv^2/2=-\mu^2/(2\lambda)$, where $(\varv/\sqrt{2})$ is the vacuum expectation value of the field $\phi$.\footnote{The vacuum expectation value 
$\varv\approx 246$~GeV is measured in muon decay thanks to its relation to the Fermi coupling $G_\mu=1/(\sqrt{2}\varv^2)$.} .
Choosing a particular field configuration that satisfies this
condition breaks the $\mathrm{SU(2)}_L \times \mathrm{U(1)}_Y$ gauge
symmetry on the vacuum and induces the spontaneous breaking
of the EW SM gauge symmetry $\mathrm{SU(2)}_L\times \mathrm{U(1)}_Y$ to a residual $\mathrm{U(1)}_{QED}$ that can be identified as the gauge symmetry of QED associated to the electric-charge quantum number $Q=I_3+Y$. In particular, by choosing
$\phi_0=(0,\varv/\sqrt{2})$, a gauge-invariant shift of the $\phi$ field
to $\phi=(0, (H+\varv)/\sqrt{2})$ allows the SM Lagrangian to display its
physical content consisting of one real scalar field, the Higgs boson
($H$), two massive (the $W^\pm_\mu$ and $Z_\mu$) and one massless (the photon
$A_\mu$) EW gauge bosons, and three families of massive fermions (quarks and charged
leptons).\footnote{This choice, corresponding to the unitary gauge, is adopted here for conciseness. For explicit calculations of radiative corrections, however, renormalizable ($R_\xi$) gauges are often more convenient since the propagators are better behaved in the ultraviolet, at the price of keeping unphysical components of the field $\phi$ explicitly.} The neutrino fields remain massless in the SM and their mass is conventionally introduced as linked to extensions of the SM that also include right-handed neutrino fields. 

The scalar field $\phi$ couples to the $W_\mu^i$ ($i=1,2,3$) and $B_\mu$ gauge fields associated with the ${\rm SU(2)}_L\times {\rm U(1)}_Y$ local
symmetry through the covariant derivative of
Eq.~(\ref{ewptests:eq:CovariantDerivative}) appearing in the kinetic
term of the scalar field $\phi$. As a result, linear combinations of
the $W^i_\mu$ ($i=1,2,3$) and $B_\mu$ fields, the $Z_\mu$ and $W^\pm_\mu$ gauge fields, acquire masses,
 \begin{equation}\label{higgs:eq:WZmasses}
M_W^2= \frac{g_2^2 \varv^2}{4}\, ,   \;\;\; M_Z^2= \frac{(g_1^2+ g_2^2) \varv^2}{4}\,,
 \end{equation} 
 and couple to the Higgs boson with strength proportional to their
 mass square as dictated by the minimal coupling induced by the
 covariant-derivative terms, namely:
\begin{equation}
\label{higgs:eq:L-Higgs-coupling-to-gauge}
{\cal{L}}_{\mathrm{Higgs}}^V =  \delta_V\frac{2M_V^2}{\varv} V_\mu V^{\mu} H+\delta_V\frac{M_V^2}{\varv^2}V_\mu V^{\mu} H^2 \,,
\end{equation}
where $V = W^{\pm}$ or $Z$ and $\delta_{W}=1, \delta_Z=1/2$. Mass and
self-couplings of the Higgs fields are defined by the $V(\phi)$
potential once expressed in terms of the Higgs-field $H$, namely:
\begin{equation}
  \label{higgs:eq:L-Higgs-mass-self-coupling}
{\cal{L}}_{\mathrm{Higgs}}^H = -\frac{1}{2}m_H^2H^2- \frac{3m_H^2}{\varv} \frac{H^3}{3!} - \frac{3m_H^2}{\varv^2} \frac{H^4}{4!}\,,
\end{equation}
where one can identify the first term as the Higgs-field mass term
($m_H^2=2\lambda \varv^2$) and the remaining two terms as the $H$-field
cubic and quartic self-interaction terms.

Finally, upon SSB the Yukawa-like interactions between the scalar field $\phi$ and
all SM massive fermions represented by the last line of
Eq.~(\ref{ewptests:eq:sm-lagrangian}) generate both the fermions' mass
terms and their couplings to the Higgs field $H$, which are therefore
related by the proportionality to the same Yukawa matrices ($\Gamma_{e,u,d}$). In
the process, both left- and right-handed fermion fields are rotated to their mass eigenstates, defined as the ones that diagonalize their mass matrix and, consequently, the corresponding Yukawa matrix.
The Yukawa Lagrangian can then be written as
\begin{equation}
  {\cal L}^f_{\mathrm{Yukawa}}=-\sum_{f^\prime}\left(m_{f^i}\bar{f}^{\prime i} f^{\prime i}
  +\frac{m_{f^i}}{\varv}\bar{f}^{\prime i} f^{\prime i}H\right)\,,
  \end{equation}
where the $f^\prime$ denote the fermion fields rotated to the mass eigenbasis. Such rotations, which differ for up-type and down-type quarks, propagate to the EW gauge currents in the second line of 
Eq.~(\ref{ewptests:eq:sm-lagrangian}) and determine the non-diagonal form of the EW charged currents, while the neutral EW currents remain flavor diagonal, as summarized in the following, dropping primes for simplicity:
\begin{eqnarray}
  {\cal L}_{\mathrm{EW-currents}}&=&
  -\sum_f  eQ_f\bar{f}^i\gamma^\mu f^iA_\mu-
  \sum_f \frac{e}{s_Wc_W}(I_3^f\bar{f}^i_L\gamma^\mu f^i_L-
  s_W^2Q_f\bar{f}^i\gamma^\mu f^i)Z_\mu -\\
  &&\frac{e}{\sqrt{2}s_W}(\bar{u}^i_L\gamma^\mu V_{ij} d^j_LW_\mu^++
  \bar{d}^i_L\gamma^\mu V^\dagger_{ij}u^j_LW_\mu^-)-
  \frac{e}{\sqrt{2}s_W}(\bar{\nu}^i_L\gamma^\mu e^i_LW_\mu^++
  \bar{e}^i_L\gamma^\mu \nu^i_LW_\mu^-)\,,\nonumber
\label{ewptests:eq:ew-currents}
\end{eqnarray}
where $f=e,\nu,u,d$ sums over all fermion fields, $V_{ij}$ are elements of the Cabibbo-Kobayashi-Maskawa matrix arising from the alluded mismatched rotation between up-type and down-type quarks, and we have
expressed all couplings in terms of the electromagnetic coupling
$e$, and the sine and cosine of the weak mixing angle $\theta_W$ using the relations:
\begin{equation}
c_W=\cos\theta_W=\frac{g_2}{\sqrt{g_1^2+g_2^2}}=\frac{M_W}{M_Z}\,,\,\,\,\, s_W=\sin\theta_W\,,\,\,\,\, e=\frac{g_1g_2}{\sqrt{g_1^2+g_2^2}}=g_1c_W=g_2s_W\,.
\label{ewptests:eq:treelevelrels}
\end{equation}

As a non-abelian spontaneously broken gauge theory, the SM is
renormalizable, and finite predictions for physical observables can be
calculated perturbatively in the EW and strong couplings, by choosing a
set of independent input parameters and a renormalization scheme that
defines them at each perturbative order. A natural choice would be to use as input parameters the parameters of the SM Lagrangian
before SSB as given in Eq.~(\ref{ewptests:eq:sm-lagrangian}). A better
choice however consists of a set of parameters that are well defined
and very precisely measured. For the purpose of EW precision tests,  besides the strong
coupling constant $\alpha_s=g_s^2/(4\pi)$,  a commonly adopted
\textit{input scheme} chooses as independent input parameters $\alpha(0)=e^2/(4\pi)$, the
electromagnetic fine structure constant in the Thomson limit
($Q^2=0$);  the Fermi constant $G_\mu$
 as measured in muon decay ($\mu^-\rightarrow e^-\bar{\nu}_e\nu_\mu$); the $Z$-boson mass
$M_Z$; the Higgs-boson mass $m_H$;
the fermion masses or equivalently their Yukawa couplings; and the
parameters of the CKM matrix $V$.\footnote{Masses are here meant to be defined in the on-shell renormalization scheme. For a more detailed discussion of the SM renormalization we refer to the corresponding chapter in this Encyclopedia.} All other EW masses and couplings
can be derived from the given set of input parameters, including all
known orders of radiative corrections. In particular, EW radiative
corrections can be absorbed into quantities commonly denoted by $\Delta r$, which summarizes EW corrections to muon decay; $\rho_Z^f$, that
modifies the vector and axial-vector coupling of the $Z$ boson to
fermion $f$,  and $\kappa_Z^f$, which collects additional corrections
to the $Z$-boson vector current, namely:
\begin{eqnarray}
 && M_W^2=\frac{M_Z^2}{2}\left(
  1+\sqrt{1-\frac{\sqrt{8}\pi\alpha(1+\Delta
    r)}{G_FM_Z^2}}\right)\,,\nonumber\\
  && g_V^f=\sqrt{\rho_Z^f}\left(
     I_3^f-2Q_f\sin^2\theta_{\mathrm{eff}}^f\right)\,,\nonumber\\
  && g_A^f=\sqrt{\rho_Z^f}I_3^f\,,\\
  && \sin^2\theta_{\mathrm{eff}}^f=\kappa_Z^f\sin^2\theta_W\,,\nonumber
  \label{eqptests:eq:ewradcorr}
\end{eqnarray}
where $g_V^f$ and $g_A^f$ are defined by rewriting the $Z$-boson
current in Eq.~(\ref{ewptests:eq:ew-currents}) in terms of vector and axial vector couplings,
i.e.
\begin{eqnarray}
  {\cal L}_{\mathrm{Z-currents}}&=&
   -\frac{e}{s_Wc_W}\sum_f (I_3^f\bar{f}^i_L\gamma^\mu f^i_L-
   s_W^2Q_f\bar{f}^i\gamma^\mu f^i)Z_\mu =
   - \frac{e}{2s_Wc_W}\sum_f \bar{f}^i(g_V^f\gamma^\mu- g_A^f\gamma^\mu\gamma^5 )f^iZ_\mu\\
   &=& -\frac{e}{2s_Wc_W}\sqrt{\rho_Z^f}\sum_f \bar{f}^i\left[
       (I_3^f-2Q_f\kappa_Z^fs_W^2)\gamma^\mu-
        I_3^f\gamma^\mu\gamma^5 )\right]f^iZ_\mu\,.\nonumber
  \end{eqnarray}

The radiative corrections to $\Delta r$ are known very precisely,
including the full one-loop EW corrections of
$O(\alpha)$~\cite{Sirlin:1980nh,Marciano:1980pb}, the full two-loop
QCD corrections of $O(\alpha\alpha_s)$~\cite{Djouadi:1987gn,
  Djouadi:1987di,Kniehl:1989yc,Halzen:1990je,Kniehl:1991gu,Kniehl:1992dx,Djouadi:1993ss}, 
three-loop QCD corrections of
$O\left(G_\mu\alpha_s^2m_t^2(1+M_Z^2/m_t^2
  +(M_Z^2/m_t^2)^2)\right)$~\cite{Avdeev:1994db,Chetyrkin:1995ix,Chetyrkin:1995js},
the full two-loop EW corrections of
$O(\alpha^2)$~\cite{Barbieri:1992nz,Barbieri:1992dq,
  Fleischer:1993ub,Fleischer:1994cb,Degrassi:1996mg,Degrassi:1996ps,Degrassi:1999jd,
  Freitas:2000gg,Freitas:2002ja,Awramik:2002wn,Onishchenko:2002ve,Awramik:2002vu,
  Awramik:2002wv,Awramik:2003ee,Awramik:2003rn}, and leading
three-loop corrections of $O(G_\mu^2\alpha_s m_t^4)$ and
$O(G_\mu^3m_t^6)$~\cite{vanderBij:2000cg,Faisst:2003px}.  The remaining theoretical
uncertainty in $M_W$ coming from missing higher-order corrections is
estimated to be 4 MeV according to Ref.~\cite{Awramik:2003rn}, where a numerical parameterization
including all known corrections is
given. This error estimate remains valid even including the recently computed higher order corrections~\cite{Boughezal:2004ef,Boughezal:2005eb,Schroder:2005db,Chetyrkin:2006bj,Boughezal:2006xk,Chen:2020xot,Chen:2020xzx}.

The radiative corrections to the effective couplings and the weak
mixing angle depend in general on the flavor of final-state fermions.
The available corrections to $\sin^2\theta_{\rm eff}^{f}$ include complete one- and two-loop EW corrections 
\cite{Marciano:1980pb,Akhundov:1985fc,Awramik:2004ge,Awramik:2006ar,Hollik:2005va,Hollik:2006ma,Awramik:2008gi,Dubovyk:2016aqv}, 
corections of $O(\alpha\alpha_s)$ to vector boson self-energies \cite{Djouadi:1987gn,Djouadi:1987di,Kniehl:1989yc,Kniehl:1991gu,Djouadi:1993ss}, non-factorizable $O(\alpha\alpha_s)$ $Zb\bar{b}$ vertex corrections \cite{Czarnecki:1996ei,Harlander:1998cmq,Fleischer:1992fq,Buchalla:1992zm,Degrassi:1993ij,Chetyrkin:1993jp}, $O(G_\mu m_t^2\alpha_s^2)$ \cite{Avdeev:1994db,Chetyrkin:1995ix}, $O(G_\mu^2 m_t^4 \alpha_s)$, $O(G_\mu^3 m_t^6)$ \cite{vanderBij:2000cg,Faisst:2003px}, and $O(G_\mu m_t^2\alpha_s^3)$ \cite{Schroder:2005db,Chetyrkin:2006bj,Boughezal:2006xk}.  A numerical parameterization including the corrections listed above was given in refs~\cite{Awramik:2004ge,Awramik:2006uz,Awramik:2008gi,Dubovyk:2019szj}. Three-loop fermionic EW and mixed EW-QCD corrections to $\sin^2\theta_W$ have been recently computed in Refs.~\cite{Chen:2020xot,Chen:2020xzx}. The corrections $\rho_Z^f$ will be discussed in Section \ref{ewptests:sec:ew-precision-observables} in the context of $Z$-boson partial decay widths.

Notice that in the limit of vanishing $g_1$ and Yukawa couplings the SM Lagrangian of Eq.~(\ref{ewptests:eq:sm-lagrangian}) displays an additional global symmetry, the so-called \textit{custodial symmetry}~\cite{Sikivie:1980hm}. In such limit, assembling the $\phi$ and $\tilde{\phi}$ SU$(2)_L$ doublets in a two-by-two matrix $\Phi = \left(\tilde{\phi}\,,\phi\right)$, the Lagrangian displays invariance under $\Phi \to U_L \Phi U_R^\dagger$, with $U_{L,R}$ independent global SU$(2)$ transformations. Hypercharge and Yukawa couplings, which distinguish right-handed $u$- and $d$-type quarks, break the custodial symmetry that also protects the relation between the $W$- and $Z$-boson masses given in Eq.~(\ref{ewptests:eq:treelevelrels}) and ensures that at tree-level $\rho\equiv M_W^2/(M_Z^2c_W^2)=1$. As a consequence, radiative corrections induced by hypercharge and Yukawa interactions break the custodial symmetry and induce deviations of $\rho$ from unity that should vanish in the limit of vanishing (or equal for up- and down-type quarks) Yukawa couplings and therefore depend quadratically on the top-quark mass, making it indeed the dominant source of custodial symmetry breaking. Also radiative corrections to $Z \to b \bar{b}$ couplings are quadratically sensitive to the top-quark mass, as can be easily seen by considering triangle diagrams with would-be Goldstone bosons and top quarks in the loop. These non-decoupling effects of the top quark can be understood since the large top-quark mass limit corresponds to large Yukawa coupling. The quadratic sensitivity to the top-quark mass has played a crucial role in the success of EW precision tests in predicting the mass range where the top quark was eventually discovered, as discussed in Section~\ref{ewptests:sec:ew-precision-fits-history}, and now strongly motivates the need for precise determinations of the top-quark mass, to fully exploit the NP sensitivity of EW precision tests.

On the other hand, the Higgs-boson mass enters EWPO only logarithmically at one-loop, and therefore the sensitivity of EW precision tests to the Higgs-boson mass is significantly milder. Still, the logarithmic sensitivity has allowed to predict a relatively light Higgs boson prior to its discovery, as discussed in Section~\ref{ewptests:sec:ew-precision-fits-history}.

\section{SM parameters and electroweak precision observables}
\label{ewptests:sec:ew-precision-observables}

The SM Lagrangian of Eq.~(\ref{ewptests:eq:sm-lagrangian}) depends on a set of input parameters, which can be chosen as discussed in Section~\ref{ewptests:sec:sm-lagrangian}. In the following we briefly discuss the current determinations of the relevant SM input parameters for EW precision tests, and then introduce the set of EWPO considered in EW precision fits, reviewing both their theoretical predictions and experimental measurements.

In the EW sector, the set of independent input parameters that are most precisely determined consists of the electromagnetic coupling constant $\alpha(0)$, the Fermi constant $G_\mu$, the $Z$-boson mass $M_Z$, and the Higgs-boson mass $m_H$. While the experimental uncertainty on $\alpha(0)$ is negligible for the purpose of EW precision tests \cite{ParticleDataGroup:2024cfk}, the relevant parameter for EWPO is the running electromagnetic coupling constant evaluated at the $Z$-boson mass scale, $\alpha(M_Z)$, which can be determined from $\alpha(0)$ through the relation:
\begin{equation}
    \alpha(M_Z) = \frac{\alpha(0)}{1-\Delta\alpha(M_Z)}\,,
\end{equation}
where $\alpha(0)=1/137.035999084(21)$ and $\Delta\alpha(M_Z)$ summarizes the contributions to the photon vacuum polarization from leptons, the five light quarks, and the top quark. While the leptonic and top-quark contributions can be calculated perturbatively with high accuracy, the contributions from the five light quarks are affected by non-perturbative QCD effects. Lattice QCD can provide a first-principle determination of the light-quark contribution to the Euclidean hadronic vacuum polarization at scales $Q^2$ below the bottom-quark mass squared, e.g.~$Q^2 = - 4$~GeV$^2$ (see Ref.~\cite{Budapest-Marseille-Wuppertal:2017okr}). The bottom-quark contribution can also be estimated on the lattice \cite{Colquhoun:2014ica}, while perturbative methods based on the Adler function can be used to determine the contributions from energy scales above the bottom-quark mass \cite{Proceedings:2019vxr}. Finally, the analytic continuation to the Minkowski space-time is performed perturbatively \cite{Proceedings:2019vxr}.\footnote{Very recently a new lattice QCD determination of the light-quark HVP together with an updated calculation of the perturbative running to $M_Z^2$ has been presented in Ref.~\cite{Conigli:2025qvh}.} An alternative approach consists of using experimental data for the cross section of $e^+e^-$ annihilation into hadrons to determine the light-quark contribution to the vacuum polarization through a dispersion relation \cite{Davier:2019can,Keshavarzi:2019abf}. 

The Fermi constant $G_\mu$ is determined with very high accuracy from the measurement of the muon lifetime $\tau_\mu$ through the relation:
\begin{equation}
    \frac{1}{\tau_\mu} = \frac{G_\mu^2 m_\mu^5}{192\pi^3} F\left(\frac{m_e^2}{m_\mu^2}\right) (1+\Delta q)\,, 
\end{equation}
where $F(x)=(1-8x+8x^3-x^4-12x^2\ln x)$ is a phase-space factor and $\Delta q$ summarizes QED radiative corrections to muon decay up to $O(\alpha^3)$ \cite{vanRitbergen:1999fi,Steinhauser:1999bx,Pak:2008qt,Fael:2020tow,Czakon:2021ybq}. The current most precise determination of $G_\mu$ is~\cite{ParticleDataGroup:2024cfk}:
\begin{equation}
    G_\mu = 1.1663787(6) \times 10^{-5}~\mathrm{GeV}^{-2}\,,
\end{equation}
with a negligible uncertainty for the purpose of EW precision tests.

The most precise determination of the $Z$-boson mass is still coming from $e^+e^-$ measurements performed at energies near the $Z$-boson resonance (LEP), and will be discussed slightly later as part of a correlated set of EWPO measured at the $Z$-boson pole.

Finally, the Higgs-boson mass has been determined with high accuracy by the ATLAS and CMS collaborations at the LHC~\cite{ATLAS:2023oaq,CMS:2024eka}. A combination of the two measurements, assuming maximal correlation between the systematic uncertainties, yields
$m_H = (125.10 \pm 0.09)~\mathrm{GeV}$.

Other input parameters relevant for EW precision tests include the strong coupling constant $\alpha_s(M_Z)$ and the mass of the top quark. The strong coupling constant can be determined from a variety of observables, including hadronic $\tau$ decays, event shapes and jet rates in $e^+e^-$ collisions, deep-inelastic scattering, and lattice QCD calculations \cite{ParticleDataGroup:2024cfk}. Numerical results presented in this chapter have been obtained by using the lattice QCD determination $\alpha_s(M_Z) =0.11873 \pm 0.00056$ from ref~\cite{Brida:2025gii}. The situation for the top-quark mass is more complex, since the quantity measured experimentally from kinematic reconstruction in hadron collider events, often denoted as the Monte Carlo (MC) top-quark mass $m_t^{MC}$, does not correspond directly to a well-defined renormalization scheme mass~\cite{Nason:2017cxd,Hoang:2020iah}. Setting aside the conceptual issues related to the definition of $m_t^{MC}$, averaging the TeVatron \cite{CDF:2016vzt}, ATLAS \cite{ATLAS:2022jbw,ATLAS:2018fwq}, and CMS \cite{CMS:2023ebf,CMS:2022kqg} measurements, and assuming maximally correlated systematic uncertainties, yields $m_t = (172.31 \pm 0.32)$~GeV.  

In terms of these input parameters, a broad set of theoretically and experimentally well known observables can be identified as optimal candidates for precision tests of the SM. The set of EWPO considered in EW precision tests includes the masses and widths of the $W$ and $Z$ bosons; the various decay rates of the $Z$ boson into leptons and quarks, including in particular the partial decay rate into $b\bar{b}$ pairs; various asymmetries measured at the $Z$ pole, including the leptonic forward-backward asymmetry, the hadronic forward-backward asymmetry for $b$ and $c$ quarks, and the left-right asymmetry measured at SLD with polarized electron beams; and finally the effective weak mixing angle as measured from various asymmetries both at LEP and SLD. These observables are all functions of the vector and axial-vector couplings of the $Z$ boson ($g_V^f$ and $g_A^f$) as defined in Eq.~(\ref{eqptests:eq:ewradcorr}), including all known SM corrections, and extracted from $Z$-pole measurements in $e^+e^-$ collisions. In the following we briefly review their definition.

The asymmetry parameter $\mathcal{A}_f$ for a given $Z\to f\bar{f}$ decay channel is 
defined in terms of the $Z$-boson effective couplings as
\begin{equation}
\mathcal{A}_f = 
\frac{2\, {\rm Re}\left(g_{V}^f/g_{A}^f\right)}
{1+\left[{\rm Re}\left(g_{V}^f/g_{A}^f\right)\right]^2}\,.
\end{equation}
The left-right asymmetry, 
the forward--backward asymmetry, and the longitudinal polarization of
the $Z\rightarrow\tau\bar{\tau}$ channel are written in terms of the asymmetry
parameters: 
\begin{eqnarray}
A_{\rm LR}^0 &=& \mathcal{A}_e\,,
\\
A_{\rm FB}^{0,f} &=& \frac{3}{4}\, \mathcal{A}_e\mathcal{A}_f\,,
\\
P_\tau^{\rm pol} &=& -\mathcal{A}_\tau\,.
\end{eqnarray}
The partial width of $Z$ decaying into a charged-lepton pair
$\ell\bar{\ell}$, including contribution from final-state QED
interactions, is given in terms of the $Z$-boson effective
couplings by~\cite{Bardin:1999ak,Bardin:1999yd}:  
\begin{align}
\Gamma_\ell &= 
\Gamma_0 \big|\rho_Z^f\big|
\sqrt{1-\frac{4m_\ell^2}{M_Z^2}}
\left[ \left(1+\frac{2m_\ell^2}{M_Z^2}\right) 
  \left(\left|\frac{g_{V}^\ell}{g_{A}^\ell}\right|^2 + 1 \right)
  - \frac{6m_\ell^2}{M_Z^2}
\right]
\left( 1 + \frac{3}{4}\frac{\alpha(M_Z^2)}{\pi}\, Q_\ell^2 \right),
\end{align}
where $\Gamma_0=G_\mu M_Z^3/(24\sqrt{2}\pi)$ and $m_\ell$ is the mass
of the final-state lepton. 
In the case of the $Z\to q\bar{q}$ channels, 
final-state QCD interactions have to be taken into account in addition
to the QED ones, and the corresponding decay rate can be written as: 
\begin{align}
\Gamma_q &= 
N_c\,
\Gamma_0 \big|\rho_Z^q\big|
\left[ \left|\frac{g_{V}^q}{g_{A}^q}\right|^2 R_V^q(M_Z^2) 
  + R_A^q(M_Z^2)
\right]
+ \Delta_{\rm EW/QCD}\,,
\end{align}
where $N_c$ is the color factor, and $R_V^q(s)$ and $R_A^q(s)$ are the
so-called radiator factors which are known to $O(\alpha_s^4)$, $O(\alpha_s \alpha)$, and $O(\alpha^2)$~\cite{Chetyrkin:1996ela,Bardin:1999ak,Bardin:1999yd,Baikov:2008jh,Baikov:2012er,Kataev:1992dg}.  The last term $\Delta_{\rm EW/QCD}$ denotes
non-factorizable EW-QCD
corrections~\cite{Bardin:1999yd,Czarnecki:1996ei,Harlander:1998cmq}.

The total decay width of the $Z$ boson, denoted by $\Gamma_Z$, is then
given by the sum of all possible channels: 
\begin{equation}
\Gamma_Z = 3\,\Gamma_\nu 
+ \Gamma_{e} + \Gamma_{\mu} + \Gamma_{\tau} + \Gamma_h\,,
\end{equation}
where we have defined the hadronic width $\Gamma_h=\sum_q \Gamma_q$. 
Moreover the ratios of widths 
\begin{eqnarray}
R_\ell^0 = \frac{\Gamma_h}{\Gamma_\ell}\,,
\qquad
R_q^0 = \frac{\Gamma_q}{\Gamma_h}\,,
\end{eqnarray}
and the cross section for $e^+e^-\to Z\to \mathrm{hadrons}$ 
at the $Z$ pole,
\begin{equation}
\sigma_h^0 =
\frac{12\pi}{M_Z^2}\frac{\Gamma_e\Gamma_h}{\Gamma_Z^2}\,,
\end{equation}
are also part of the EWPO. 

The Z-boson partial and total decay widths as well as the branching ratios and the hadronic cross-section are known including one- and two-loop EW corrections~\cite{Akhundov:1985fc,Freitas:2013dpa,Freitas:2014hra,Dubovyk:2018rlg}, corrections of $O(\alpha\alpha_s)$ to vector boson self-energies~\cite{Djouadi:1987gn,Djouadi:1987di,Kniehl:1989yc,Kniehl:1991gu,Djouadi:1993ss}, non-factorizable $O(\alpha\alpha_s)$ $Zq\bar{q}$ vertex corrections~\cite{Czarnecki:1996ei,Harlander:1998cmq,Fleischer:1992fq,Buchalla:1992zm,Degrassi:1993ij,Chetyrkin:1993jp}, $O(G_\mu m_t^2\alpha_s^2)$~\cite{Avdeev:1994db,Chetyrkin:1995ix}, $O(G_\mu^2 m_t^4 \alpha_s)$, $O(G_\mu^3 m_t^6)$~\cite{vanderBij:2000cg,Faisst:2003px}, and $O(G_\mu m_t^2\alpha_s^3)$~\cite{Schroder:2005db,Chetyrkin:2006bj,Boughezal:2006xk} corrections. Three-loop fermionic EW and mixed EW-QCD corrections to decay rates have been recently computed in Refs.~\cite{Chen:2020xot,Chen:2020xzx}. 

The $W$-boson decay width $\Gamma_W$ is known at one loop~\cite{Bardin:1999ak,Bardin:1986fi,Bardin:1981sv,Denner:1990cpz,Denner:1991kt,Kniehl:2000rb,Chang:1981qq}, plus corrections of $O(\alpha \alpha_s)$~\cite{Kara:2013dua} and QCD corrections up to $O(\alpha_s^4)$~\cite{Gorishnii:1990vf,Baikov:2008jh,Baikov:2012er}.

On the experimental side, the Z-pole observables have been measured with high accuracy at LEP and SLD. From the experimental measurements the pseudo-observables defined above have been extracted and combined by the LEP and SLD Electroweak working groups~\cite{ALEPH:2005ab}. Under the hypothesis of lepton universality, the pseudo-observables measured at the $Z$ pole can be divided into two correlated sets: the first set includes $M_Z$, $\Gamma_Z$, $\sigma_h^0$, $R_\ell^0$, $A_{\rm FB}^{0,\ell}$, while the second set includes $R_b^0$, $R_c^0$, $A_{\rm FB}^{0,b}$, $A_{\rm FB}^{0,c}$, $\mathcal{A}_\ell$, $\mathcal{A}_b$, and $\mathcal{A}_c$. The first set has been updated in Ref.~\cite{Janot:2019oyi} by reanalyzing the LEP data with improved calculations of the Bhabha-scattering cross section used for the luminosity determination. In the second set, $A^{0,b}_{\rm FB}$
has been updated
by including $b$-quark mass effects in the NNLO QCD
corrections~\cite{Bernreuther:2016ccf} \footnote{The original measurements have been recently reanalyzed with modern Monte Carlo
parton-shower tools in Ref.~\cite{dEnterria:2020cgt}, leading to a reduction of the subdominant QCD uncertainty in the measurement, without changing the total experimental uncertainty.},
relevant for the extraction of the pseudo-observable from data. The experimental values are reported in Table~\ref{ewptests:tab:SMfit}, where the correlated sets are highlighted in different shades of gray patterns. Further uncorrelated measurements at the $Z$ pole include $\sin^2\theta_\mathrm{eff}^\mathrm{lept}$ as extracted from the hadronic forward-backward charge asymmetry $Q_\mathrm{FB}^\mathrm{had}$, $P_\tau^\mathrm{pol} \equiv -\mathcal{A}_\ell$, and the $s$-quark asymmetry parameter $\mathcal{A}_s$ as measured by SLD \cite{Abe:2000uc}.

While the Z-pole observables have been measured at $e^+e^-$ colliders, the $W$-boson mass and width have been measured both at LEP2 \cite{LEP-2} and at hadron colliders (Tevatron \cite{TevatronElectroweakWorkingGroup:2010mao,D0:2012kms,CDF:2022hxs} and LHC \cite{LHCb:2021bjt,ATLAS:2024erm,CMS:2024lrd}). Recently, the CDF collaboration at Tevatron has presented a new measurement of the $W$-boson mass with an accuracy of 9 MeV \cite{CDF:2022hxs}, which was about a factor of two better than the previous world average. However, the CDF measurement showed a significant tension with both the SM prediction and previous measurements performed at LEP2 and at the Tevatron by the D0 collaboration.  This tension revamped the interest in EW precision tests as an extremely sensitive probe of NP effects. However, the tension between the different experimental determinations called for a careful attitude in averaging the various measurements. As discussed in Ref.~\cite{deBlas:2022hdk}, the tension generated in the EW fit by the CDF measurement could be significantly alleviated if performing a PDG-style average with a suitable scale factor to account for the existing tensions among the different measurements. At the same time, LHC experiments have since then performed new measurements of the $W$-boson mass. In particular, the very recent CMS measurement \cite{CMS:2024lrd} has an accuracy comparable to the CDF one, and is in good agreement with the previous world average excluding CDF. Therefore, in the numerical results presented in this chapter we have chosen to use the average of all measurements excluding the CDF one.  The average is performed assuming a correlated systematic PDF uncertainty between the LHC measurements, yielding $M_W = (80.3635 \pm 0.0080)$~GeV.

Hadron colliders have provided also precise measurements of the effective leptonic weak mixing angle, $\sin^2\theta_{\mathrm{eff}}^{\mathrm{lept}}$. Available measurements include the Tevatron
determinations in Ref.~\cite{Aaltonen:2018dxj}, the LHCb
measurement in Refs.~\cite{Aaij:2015lka,LHCb:2024ygc}, as well as the ATLAS~\cite{Aad:2015uau,ATLAS:2018gqq} and
CMS~\cite{Sirunyan:2018swq,CMS:2024ony} measurements. The combined value used in the numerical analyses presented in this Chapter is $\sin^2\theta_{\mathrm{eff}}^{\mathrm{lept}} = (0.23150 \pm 0.00023)$. CMS also measured $P_\tau^{\mathrm{pol}}$ in Ref.~\cite{CMS:2023mgq}.

Measurements of the $W$ branching ratios can also be included, both from LEP2~\cite{ALEPH:2013dgf} and from the recent (and more precise) determination by CMS~\cite{CMS:2022mhs}.

\section{Electroweak precision fits}
\label{ewptests:sec:ew-precision-fits}
As discussed in the introductory overview of Section~\ref{ewpf:sec:intro},
several relations among particle masses and couplings are predicted in the SM as a consequence of the underlying gauge symmetry and its spontaneous breaking. These relations can be tested with high precision by comparing theoretical predictions and experimental measurements of the set of EWPO presented in Section~\ref{ewptests:sec:ew-precision-observables}. The comparison is most efficiently performed through a global fit that takes into account both the theoretical correlations among different observables as induced by their dependence on a common set of input parameters, and the experimental correlations of their measurements. 
%In this Section we will review the framework used to perform EW precision fits, highlight the crucial role played by EW precision fits of the SM in predicting the mass of the top quark and of the Higgs boson prior to their discovery, present an updated EW precision fit of the SM, and illustrate how EW precision fits can give general constraints on physics beyond the SM.

%In this Section we will review the framework used to perform EW
%precision fits, highlight the crucial role played by the electroweak
%fit in predicting the mass of the top quark and of the Higgs boson
%prior to their discovery, present an updated EW precision fit of the
%SM, and illustrate how EW precision fits can give general constraints
%on physics beyond the SM.
In the following, after briefly recalling in Section~\ref{ewptests:sec:ew-precision-fits-history} the crucial
role played by SM EW precision fits in predicting the mass of the top
quark and of the Higgs boson prior to their discovery, we will focus
on the current status of EW precision fits of the SM. We will introduce their
general framework in Section~\ref{ewptests:sec:ew-precision-fits-general} and present state-of-the-art
results in Section~\ref{ewptests:sec:ew-precision-fits-sm}. 

\subsection{Historical role of EW precision fits: prediction of \texorpdfstring{$m_t$ and $m_H$}{mt and mH}}
\label{ewptests:sec:ew-precision-fits-history}

\begin{figure}[!h]
  \centering
  \begin{minipage}[t]{0.45\textwidth}
      \includegraphics[height=.25\textheight]{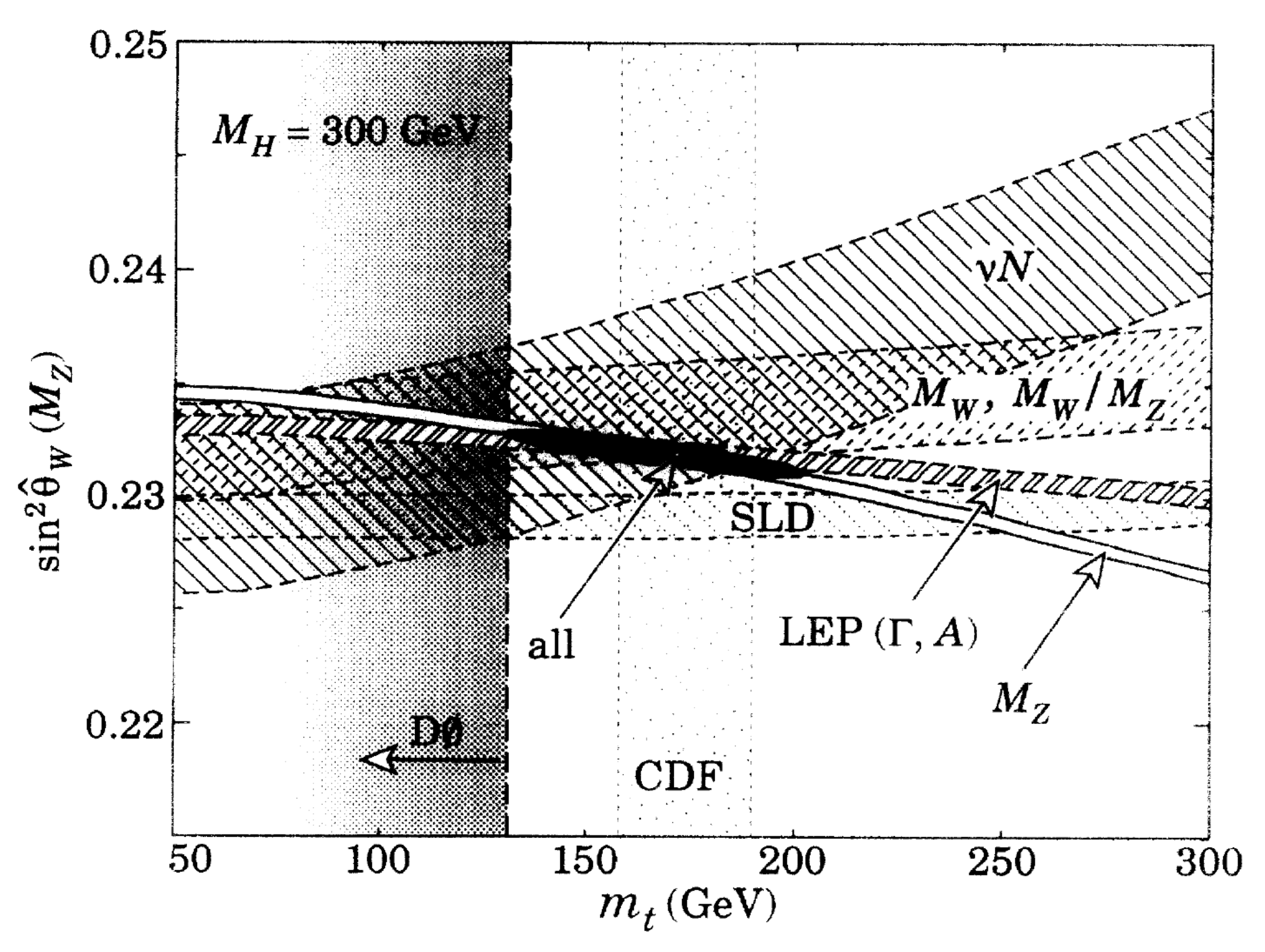}
  \end{minipage}
  \hspace{0.5truecm}
  \begin{minipage}[b]{0.5\textwidth}{
      \includegraphics[height=.27\textheight]{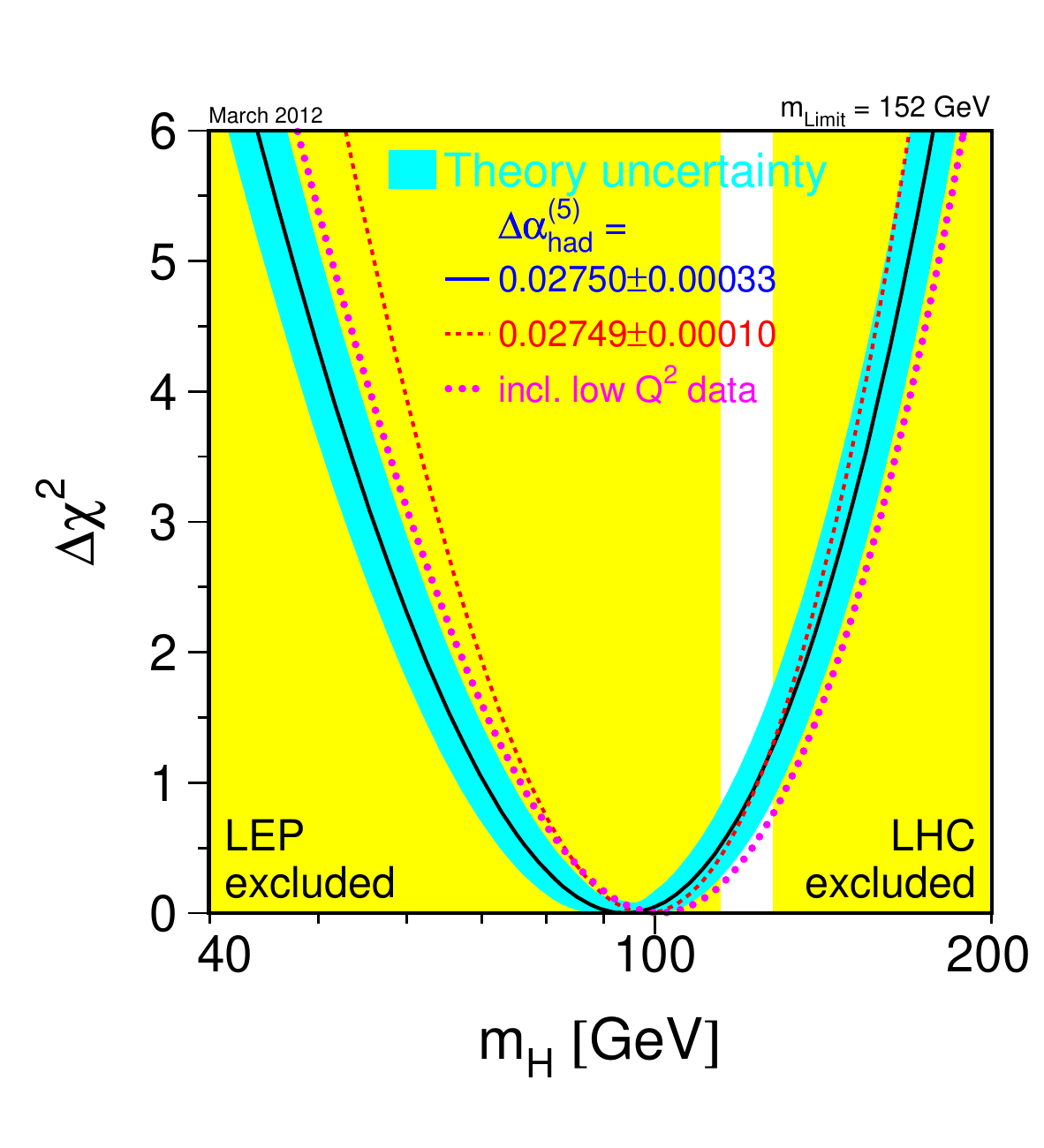}}
  \end{minipage}
  \caption{Left panel: Impact of different EWPO (neutrino-nucleon scattering, $M_Z$ and $M_W$, $Z$-decay rates and asymmetries at LEP) on the indirect determination of $m_t$ and $\sin^2\theta$ from the 1994 PDG review \cite{ParticleDataGroup:1994kdp}, assuming $m_H=300$ GeV, together with the direct determinations of $m_t$ (pre-discovery range for CDF, lower bound for D0) and $\sin^2 \theta$ from SLD. Right panel: Indirect determination of $m_H$ from the EW precision fit as a function of the uncertainty on $\Delta \alpha^{(5)}_\mathrm{had}$, together with the LEP lower bound and the LHC upper bound, presented by the LEP-TeVatron Electroweak Working Group in 2012 \cite{LEPEWWG}.}
  \label{fig:oldEW}
\end{figure}

Electroweak precision fits have played a crucial role in the history
of particle physics by predicting the mass of the top quark and of the
Higgs boson prior to their discovery. In the early 1990s, before the
discovery of the top quark at the Tevatron collider in 1995, the
Particle Data Group (PDG) review on EW physics in the SM
\cite{ParticleDataGroup:1994kdp} presented an indirect determination
of the weak mixing angle, the top-quark mass and the strong coupling
constant. Assuming a Higgs-boson mass of $m_H=300$ GeV, the fit
yielded $m_t=(169^{+16}_{-18})$ GeV (see Figure~\ref{fig:oldEW}, left
panel), while varying $m_H$ in the range $[60,1000]$ GeV changed the
central value of $m_t$ by $+17/-20$~GeV. This prediction was in
remarkable agreement with the direct measurement of $m_t=(180\pm 12)$
GeV obtained at the Tevatron shortly after
\cite{CDF:1995wbb,D0:1995jca}. Having directly measured the top-quark
mass, the EW precision fit was then used to predict the mass of the
Higgs boson, although with a larger uncertainty due to the logarithmic
dependence of the EWPO on $m_H$. Right before the discovery of the
Higgs boson at the LHC in 2012, the LEP-Tevatron Electroweak Working Group presented~\cite{LEPEWWG} an indirect determination of the Higgs-boson mass based on the EW precision fit (see Figure~\ref{fig:oldEW}, right panel), yielding $m_H=(94^{+29}_{-24})$ GeV with an upper bound of $154$ GeV, in good agreement with the direct measurement of $m_H=(125.09\pm 0.24)$ GeV son after obtained at the LHC~\cite{Aad:2012tfa,Chatrchyan:2012xdj}.

\subsection{General framework}
\label{ewptests:sec:ew-precision-fits-general}

After the discovery of the Higgs boson, EW precision fits of the SM
have focused on testing the consistency of the theory to the highest
available accuracy. Different frameworks mainly differ by the adopted
statistical approach and occasionally by the choice of EWPO or the
assumptions made in combining available experimental measurements.
The results presented in
Section~\ref{ewptests:sec:ew-precision-fits-sm} are based on the
theoretical predictions and experimental measurements described in
Section~\ref{ewptests:sec:ew-precision-observables}.  

For the sake of concreteness, we adopt the Bayesian framework implemented in the \texttt{HEPfit} package~\cite{deBlas:2019ehy}; similar results can be obtained in a frequentist framework~\cite{Haller:2018nnx,Erler:2019hds}. In the Bayesian approach, the starting point is Bayes' theorem, which relates the posterior probability distribution $p(\theta|D,M)$ of a set of model parameters $\theta$ given a set of data $D$ and a model $M$, to the prior probability distribution $\pi(\theta|M)$ of the parameters in the model and the likelihood function $\mathcal{L}(D|\theta,M)$ of the data given the parameters in the model, as follows:
\begin{equation}
    p(\theta|D,M) = \frac{\mathcal{L}(D|\theta,M)\pi(\theta|M)}{p(D|M)}\,,
\end{equation}
where $p(D|M)$ is the evidence or marginal likelihood, which plays the role of a normalization factor. In the case of EW precision fits of the SM, the model $M$ is the SM itself, the parameters $\theta$ are the SM input parameters relevant for the calculation of the EWPO (see Section~\ref{ewptests:sec:sm-lagrangian}), and the data $D$ are the experimental measurements of the EWPO (see Section~\ref{ewptests:sec:ew-precision-observables}). The likelihood function $\mathcal{L}(D|\theta,M)$ is constructed as a multi-dimensional Gaussian distribution that takes into account both experimental and theoretical uncertainties as well as correlations among different EWPO. Note that the knowledge of the normalization factor $p(D|M)$ is not required in order to determine probability intervals for the parameters or the observables, since they do not require the posterior distribution to be normalized. From the fit one obtains also posterior distributions for the EWPO. 

First of all, prior distributions for the selected set of SM input parameters are chosen to be uniform distributions in a range that is
sufficiently broad to encompass the corresponding posterior distributions.
%First of all, we define prior distributions for the SM input
%parameters discussed in Section~\ref{ewptests:sec:sm-lagrangian} (we
%stick to the $\alpha$ scheme for convenience). These are chosen to be uniform distributions in a range that is sufficiently broad to encompass the posterior distribution. 
Having defined the input-parameter priors, there are several types of analyses that can be performed:
\begin{itemize}
    \item \textit{Full fit}: all available information on all input parameters and EWPO is used to determine the posterior distribution of all input parameters and EWPO. This represents our best knowledge of the SM parameters and of the EWPO within the SM.
    \item \textit{Individual Prediction}: information on a single EWPO is removed from the full fit, and the \emph{posterior predictive distribution} of that EWPO is determined based on the information on all input parameters and all other EWPO. For EWPO that have an experimental determination which is uncorrelated with other EWPO, the consistency of the SM can be quantified by computing the \textit{pull} between the two determinations and the corresponding \emph{p-value}. To this aim, we sample the posterior predictive distribution and the experimental one, construct the probability density function (p.d.f.) $p(x)$ of the residuals $x\equiv (x_\mathrm{exp} - x_\mathrm{th})$, and compute the integral of the p.d.f. in the region $p(x) < p(0)$. This two-sided p-value is then converted to the equivalent number of standard deviations for a Gaussian distribution, which we quote as 1D Pull in Table \ref{ewptests:tab:SMfit}. In the case of a Gaussian posterior predictive distribution, this quantity coincides with the
usual pull defined as the difference between the central values of
the two distributions divided by the sum in quadrature of the
residual mean square of the distributions themselves.
The advantage of this approach is that no approximation is made
on the shape of p.d.f.'s. 
For experimentally correlated EWPO, one can also
define an \emph{nD pull}
by removing from the fit one set of correlated observables at a time and
computing the prediction for the set of observables together with their
covariance matrix. Then the same procedure described for \emph{1D pulls} can be carried out, this time sampling the posterior predictive and experimental n-dimensional p.d.f.'s. This {\emph{nD pull}} is shown in
the last column in Table~\ref{ewptests:tab:SMfit}. 
    \item \textit{Individual Indirect determination}: the information on a single input parameter is removed from the full fit and the posterior predictive distribution of that input parameter is determined based on the information on all other input parameters and all EWPO. The pull and $p$-value between this indirect determination and the information on that input parameter used in the full fit can be computed as described above for the \emph{Individual Prediction} case.
    \item \textit{Full Prediction}: only the information on the SM input parameters is used to determine the posterior predictive distribution of all EWPO. This represents the SM prediction for the EWPO based purely on the knowledge of the input parameters. The posterior predictive distribution can be directly compared with the experimental measurements of the EWPO, computing the pull and $p$-value between the two determinations as described above for the \emph{Individual Prediction} case.
    \item \textit{Full Indirect determination}: only the information on the EWPO is used to determine the posterior predictive distribution of all input parameters. This represents the indirect determination of the SM input parameters based purely on the knowledge of the EWPO. Also in this case the pull and $p$-value can be defined and computed as described above.
\end{itemize}
In Section~\ref{ewptests:sec:ew-precision-fits-sm} we will present results for all the different fits and discuss them in more detail.

\subsection{Updated EW precision fits of the SM}
\label{ewptests:sec:ew-precision-fits-sm}
\begin{table}[!b]
\centering
\scriptsize
\begin{tabular}{lllllll}
  \toprule
&  \multicolumn{1}{c}{Full Prediction} & \multicolumn{1}{c}{Total parametric} &
${\alpha_s}(M_{Z}^2)$ &
${\Delta\alpha_{\rm had}^{(5)}}(M_{Z}^2)$ &
\quad ${M_Z}$ &
\quad ${m_t}$ \\
  \midrule
$M_W$ [GeV] & $ 80.3524 \pm 0.0055 $
& \quad $\pm 0.0038 $ % _total
& $\pm 0.0001 $ % _AlsMz
& $\pm 0.0018 $ % _dAle5Mz
& $\pm 0.0027 $ % _Mz
& $\pm 0.0019 $ % _mtop
\\ 
$\Gamma_{W}$ [GeV] & $ 2.08811 \pm 0.00048 $
& \quad $\pm 0.00037 $ % _total
& $\pm 0.00024 $ % _AlsMz
& $\pm 0.00014 $ % _dAle5Mz
& $\pm 0.00021 $ % _Mz
& $\pm 0.00015 $ % _mtop
\\ 
$\Gamma_{Z}$ [GeV] & $ 2.49457 \pm 0.00055 $
& \quad $\pm 0.00037 $ % _total
& $\pm 0.00029 $ % _AlsMz
& $\pm 0.00010 $ % _dAle5Mz
& $\pm 0.00021 $ % _Mz
& $\pm 0.00007 $ % _mtop
\\ 
$\sigma_{h}^{0}$ [nb] & $ 41.4875 \pm 0.0069 $
& \quad $\pm 0.0034 $ % _total
& $\pm 0.0027 $ % _AlsMz
& $\pm 0.0001 $ % _dAle5Mz
& $\pm 0.0020 $ % _Mz
& $\pm 0.0002 $ % _mtop
\\ 
$\sin^2\theta_{\rm eff}^{\rm lept}$ & $ 0.231545 \pm 0.000061 $
& \quad $\pm 0.000039 $ % _total
& $\pm 0.000000 $ % _AlsMz
& $\pm 0.000035 $ % _dAle5Mz
& $\pm 0.000015 $ % _Mz
& $\pm 0.0000010 $ % _mtop
\\ 
$\mathcal{A}_{\ell}$ & $ 0.14684 \pm 0.00048 $
& \quad $\pm 0.00031 $ % _total
& $\pm 0.00000 $ % _AlsMz
& $\pm 0.00027 $ % _dAle5Mz
& $\pm 0.00012 $ % _Mz
& $\pm 0.00008 $ % _mtop
\\ 
$\mathcal{A}_c$ & $ 0.66771 \pm 0.00022 $
& \quad $\pm 0.00014 $ % _total
& $\pm 0.00000 $ % _AlsMz
& $\pm 0.00012 $ % _dAle5Mz
& $\pm 0.00005 $ % _Mz
& $\pm 0.00004 $ % _mtop
\\ 
$\mathcal{A}_b$ & $ 0.934724 \pm 0.000040 $
& \quad $\pm 0.000025 $ % _total
& $\pm 0.000000 $ % _AlsMz
& $\pm 0.000023 $ % _dAle5Mz
& $\pm 0.000010 $ % _Mz
& $\pm 0.000002 $ % _mtop
\\ 
$A_{\rm FB}^{0, \ell}$ & $ 0.01617 \pm 0.00011 $
& \quad $\pm 0.000068 $ % _total
& $\pm 0.000001 $ % _AlsMz
& $\pm 0.000060 $ % _dAle5Mz
& $\pm 0.000026 $ % _Mz
& $\pm 0.000017 $ % _mtop
\\ 
$A_{\rm FB}^{0, c}$ & $ 0.07354 \pm 0.00025 $
& \quad $\pm 0.00017 $ % _total
& $\pm 0.00000 $ % _AlsMz
& $\pm 0.00015 $ % _dAle5Mz
& $\pm 0.00006 $ % _Mz
& $\pm 0.00004 $ % _mtop
\\ 
$A_{\rm FB}^{0, b}$ & $ 0.10295 \pm 0.00034 $
& \quad $\pm 0.00022 $ % _total
& $\pm 0.00000 $ % _AlsMz
& $\pm 0.00020 $ % _dAle5Mz
& $\pm 0.00008 $ % _Mz
& $\pm 0.00005 $ % _mtop
\\ 
$R^{0}_{\ell}$ & $ 20.7531 \pm 0.0070 $
& \quad $\pm 0.0035 $ % _total
& $\pm 0.0034 $ % _AlsMz
& $\pm 0.0006 $ % _dAle5Mz
& $\pm 0.0003 $ % _Mz
& $\pm 0.0001 $ % _mtop
\\ 
$R^{0}_{c}$ & $ 0.172215 \pm 0.000051 $
& \quad $\pm 0.000012 $ % _total
& $\pm 0.000012 $ % _AlsMz
& $\pm 0.000002 $ % _dAle5Mz
& $\pm 0.000001 $ % _Mz
& $\pm 0.000004 $ % _mtop
\\ 
$R^{0}_{b}$ & $ 0.21588 \pm 0.00010 $
& \quad $\pm 0.000012 $ % _total
& $\pm 0.000008 $ % _AlsMz
& $\pm 0.000001 $ % _dAle5Mz
& $\pm 0.000000 $ % _Mz
& $\pm 0.000011 $ % _mtop
\\ 
BR$_{W_{\ell \nu}}$ & $ 0.108362 \pm 0.000013 $
& \quad $\pm 0.000013 $ % _total
& $\pm 0.000013 $ % _AlsMz
& $\pm 0.000000 $ % _dAle5Mz
& $\pm 0.000000 $ % _Mz
& $\pm 0.000000 $ % _mtop
\\ 
$A_{s}$ & $ 0.935631 \pm 0.000040 $
& \quad $\pm 0.000025 $ % _total
& $\pm 0.000001 $ % _AlsMz
& $\pm 0.000022 $ % _dAle5Mz
& $\pm 0.000010 $ % _Mz
& $\pm 0.000006 $ % _mtop
\\ 
\bottomrule
\end{tabular}
\caption{Full prediction, total parametric uncertainties and individual contributions to the parametric uncertainty of each SM parameter, except for $m_H$ (see text). Individual contributions are obtained setting all SM parameters to their central values, except for the one indicated in each column, which is allowed to float according to its uncertainty. Individual parametric uncertainties are rounded to the second significant digit of the total one.\label{ewptests:tab:SMpred}}
\end{table}

As described in Section~\ref{ewptests:sec:ew-precision-fits-general},
the results presented on the following have been obtained assuming
flat priors for all the SM input parameters, while including the
information of their experimental measurements in the likelihood.
{\it Intrinsic} theoretical uncertainties due to missing
higher-order corrections to EWPO are also included in the fits,\footnote{These uncertainties are implemented in the fit as nuisance parameters with Gaussian prior distributions.} using the results of Ref.~\cite{Dubovyk:2018rlg} to which we refer for more
details. The main theory uncertainties considered are:
\begin{eqnarray}
    &&\delta_\mathrm{th} M_W = 4\, \mathrm{MeV}\,,\quad \delta_\mathrm{th} \sin^2{\theta_W} = 5\cdot 10^{-5}\,,\quad \delta_\mathrm{th} \Gamma_{Z} = 0.4\, \mathrm{MeV} \,,\quad
    \delta_\mathrm{th} \sigma^0_\mathrm{had} = 6\, \mathrm{pb} \,,\quad \nonumber \\
    &&\delta_\mathrm{th} R^0_\ell = 0.006 \,,\quad \delta_\mathrm{th} R^0_c = 0.00005 \,,\quad
    \delta_\mathrm{th} R^0_b = 0.0001 \,.\quad
    \label{eq:therr}
\end{eqnarray}
Theoretical uncertainties are
still small compared to the experimental ones and, therefore, they have a very
small impact on the fit. The same applies to the {\it parametric}
theory uncertainties, obtained by propagating the experimental errors
of the SM input parameters into the predictions for the EWPO.  The breakdown of
these parametric errors is detailed in Table~\ref{ewptests:tab:SMpred}, except
for the contributions coming from the uncertainty in $m_H$, which are numerically irrelevant in the total parametric uncertainty. Besides the \emph{Full Prediction}, the second column in Table~\ref{ewptests:tab:SMpred} shows the total parametric uncertainty, while the remaining columns show the breakdown of the parametric uncertainty into the contributions from ${\alpha_s}(M_{Z}^2)$, ${\Delta\alpha_{\rm had}^{(5)}}(M_{Z}^2)$, ${M_Z}$, and ${m_t}$.

\begin{figure}[!t]
  \centering
  \includegraphics[height=.7\textheight]{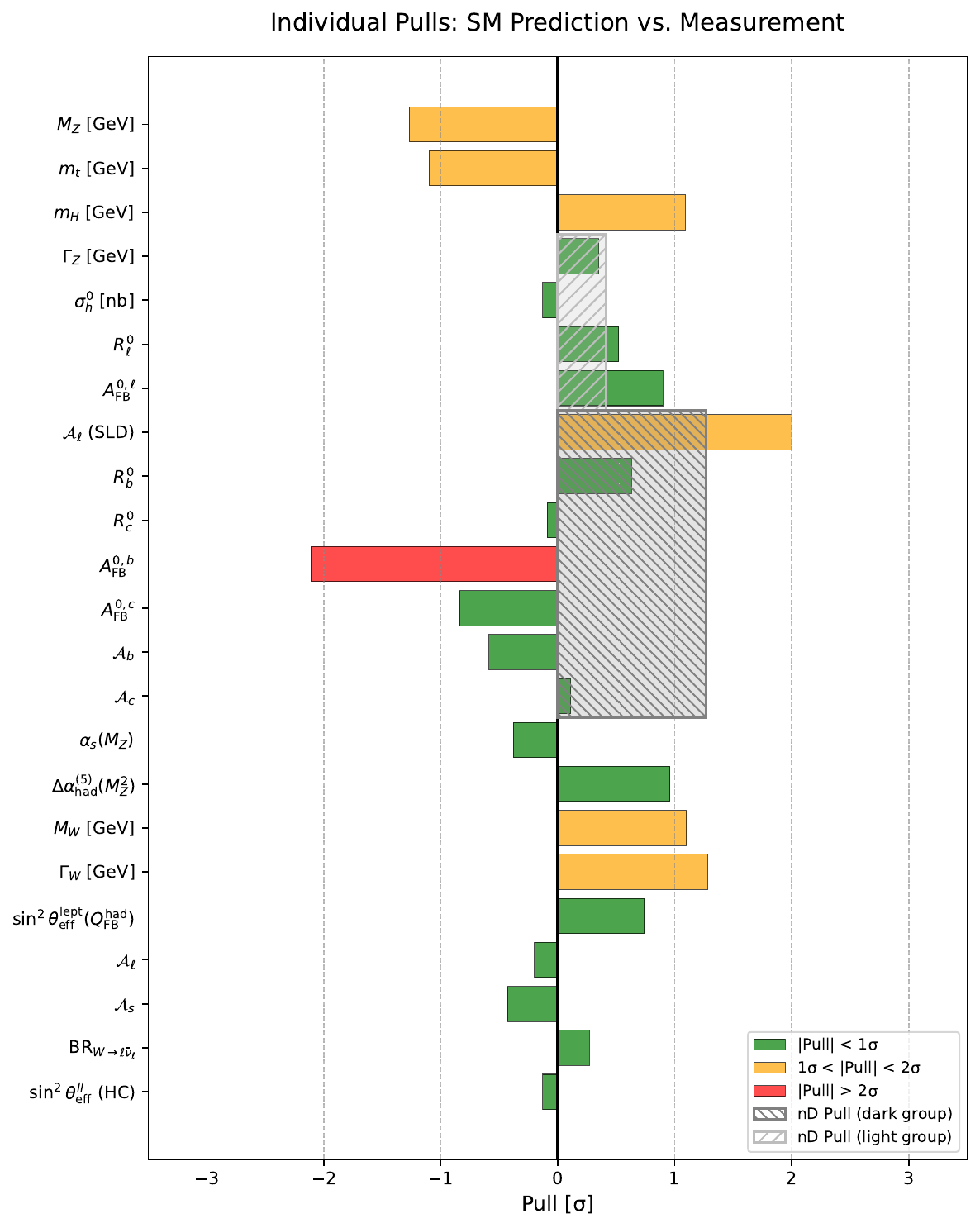}
  \caption{Summary of the pulls between the experimental measurements and the individual predictions from the global SM EW fit, as reported in Table~\ref{ewptests:tab:SMfit}.}
  \label{ewptests:fig:pulls}
\end{figure}

\begin{table}[!htb]
{
\begin{center}
  \scriptsize
\begin{tabular}{lccccc}
\toprule
\multicolumn{6}{c}{Global SM EW fit } \\
&  &  &   &  &  \\ [-0.2cm]
& Measurement & Posterior & Individual Prediction &1D Pull &nD Pull \\
\hline 
\hline 
$\alpha_{s}(M_{Z})$ & $ 0.11873 \pm 0.00056 $ & $ 0.11878 \pm 0.00055 $  & $ 0.1199 \pm 0.0028 $& -0.38 &  \\ 
& & $[ 0.11769 , 0.11986 ]$ & $[ 0.1144 , 0.1253 ]$ & &  \\ 
$\Delta\alpha^{5}_{\mathrm{had}}(M_{Z}^{2})$ & $ 0.02766 \pm 0.00010 $ & $ 0.027631 \pm 0.000096 $  & $ 0.02734 \pm 0.00032 $& 0.96 &  \\ 
& & $[ 0.027443 , 0.027819 ]$ & $[ 0.02670 , 0.02797 ]$ & &  \\ 
\rowcolor[gray]{.8} $M_Z$ [GeV] & $ 91.1875 \pm 0.0021 $ & $ 91.1885 \pm 0.0019 $  & $ 91.1964 \pm 0.0068 $& -1.27 &  \\ 
\rowcolor[gray]{.8} & & $[ 91.1847 , 91.1923 ]$ & $[ 91.1832 , 91.2097 ]$ & &  \\ 
$m_t$ [GeV] & $ 172.31 \pm 0.32 $ & $ 172.38 \pm 0.31 $  & $ 174.0 \pm 1.5 $& -1.10 &  \\ 
& & $[ 171.77 , 173.00 ]$ & $[ 171.1 , 177.0 ]$ & &  \\ 
$m_H$ [GeV] & $ 125.100 \pm 0.090 $ & $ 125.101 \pm 0.090 $  & $ 106.2 \pm 16.3 $& 1.09 &  \\ 
& & $[ 124.925 , 125.278 ]$ & $[ 76.2 , 141.4 ]$ & &  \\ 
\hline 
$M_W$ [GeV] & $ 80.3635 \pm 0.0080 $ & $ 80.3562 \pm 0.0045 $  & $ 80.3529 \pm 0.0054 $& 1.10 &  \\ 
& & $[ 80.3475 , 80.3650 ]$ & $[ 80.3423 , 80.3635 ]$ & &  \\ 
\hline 
$\Gamma_{W}$ [GeV] & $ 2.151 \pm 0.049 $ & $ 2.08844 \pm 0.00041 $  & $ 2.08843 \pm 0.00042 $& 1.28 &  \\ 
& & $[ 2.08762 , 2.08925 ]$ & $[ 2.08761 , 2.08923 ]$ & &  \\ 
\hline 
$\sin^2\theta_{\rm eff}^{\rm lept}(Q_{\rm FB}^{\rm had})$ & $ 0.2324 \pm 0.0012 $ & $ 0.231523 \pm 0.000055 $  & $ 0.231523 \pm 0.000057 $& 0.74 &  \\ 
& & $[ 0.231415 , 0.231631 ]$ & $[ 0.231412 , 0.231634 ]$ & &  \\ 
\hline 
$-P_{\tau}^{\rm pol}=\mathcal{A}_\ell$ & $ 0.1465 \pm 0.0033 $ & $ 0.14701 \pm 0.00043 $  & $ 0.14702 \pm 0.00043 $& -0.20 &  \\ 
& & $[ 0.14616 , 0.14785 ]$ & $[ 0.14616 , 0.14788 ]$ & &  \\ 
\hline 
\rowcolor[gray]{.8} $\Gamma_{Z}$ [GeV] & $ 2.4955 \pm 0.0023 $ & $ 2.49474 \pm 0.00052 $  & $ 2.49466 \pm 0.00054 $& 0.35 &  \\ 
\rowcolor[gray]{.8} & & $[ 2.49372 , 2.49577 ]$ & $[ 2.49361 , 2.49573 ]$ & &  \\ 
\rowcolor[gray]{.8} $\sigma_{h}^{0}$ [nb] & $ 41.480 \pm 0.033 $ & $ 41.4862 \pm 0.0067 $  & $ 41.4871 \pm 0.0068 $& -0.13 & 0.41 \\ 
\rowcolor[gray]{.8} & & $[ 41.4731 , 41.4995 ]$ & $[ 41.4735 , 41.5005 ]$ & &  \\ 
\rowcolor[gray]{.8} $R^{0}_{\ell}$ & $ 20.767 \pm 0.025 $ & $ 20.7545 \pm 0.0067 $  & $ 20.7529 \pm 0.0069 $& 0.52 &  \\ 
\rowcolor[gray]{.8} & & $[ 20.7413 , 20.7676 ]$ & $[ 20.7394 , 20.7668 ]$ & &  \\ 
\rowcolor[gray]{.8} $A_{\rm FB}^{0, \ell}$ & $ 0.0171 \pm 0.0010 $ & $ 0.016208 \pm 0.000095 $  & $ 0.016197 \pm 0.000096 $& 0.90 &  \\ 
\rowcolor[gray]{.8} & & $[ 0.016023 , 0.016397 ]$ & $[ 0.016011 , 0.016386 ]$ & &  \\ 
\hline 
\rowcolor[gray]{.7} $\mathcal{A}_{\ell}$ (SLD) & $ 0.1513 \pm 0.0021 $ & $ 0.14701 \pm 0.00043 $  & $ 0.14700 \pm 0.00045 $& 2.00 &  \\ 
\rowcolor[gray]{.7} & & $[ 0.14616 , 0.14785 ]$ & $[ 0.14613 , 0.14789 ]$ & &  \\ 
\rowcolor[gray]{.7} $R^{0}_{b}$ & $ 0.21629 \pm 0.00066 $ & $ 0.215881 \pm 0.0000100 $  & $ 0.21588 \pm 0.00010 $& 0.63 &  \\ 
\rowcolor[gray]{.7} & & $[ 0.215684 , 0.216074 ]$ & $[ 0.21567 , 0.21607 ]$ & &  \\ 
\rowcolor[gray]{.7} $R^{0}_{c}$ & $ 0.1721 \pm 0.0030 $ & $ 0.172218 \pm 0.000052 $  & $ 0.172218 \pm 0.000051 $& -0.09 &  \\ 
\rowcolor[gray]{.7} & & $[ 0.172117 , 0.172320 ]$ & $[ 0.172116 , 0.172319 ]$ & &  \\ 
\rowcolor[gray]{.7} $A_{\rm FB}^{0, b}$ & $ 0.0996 \pm 0.0016 $ & $ 0.10306 \pm 0.00031 $  & $ 0.10305 \pm 0.00031 $& -2.11 & 1.27 \\ 
\rowcolor[gray]{.7} & & $[ 0.10247 , 0.10366 ]$ & $[ 0.10244 , 0.10368 ]$ & &  \\ 
\rowcolor[gray]{.7} $A_{\rm FB}^{0, c}$ & $ 0.0707 \pm 0.0035 $ & $ 0.07363 \pm 0.00022 $  & $ 0.07362 \pm 0.00023 $& -0.84 &  \\ 
\rowcolor[gray]{.7} & & $[ 0.07318 , 0.07407 ]$ & $[ 0.07316 , 0.07409 ]$ & &  \\ 
\rowcolor[gray]{.7} $\mathcal{A}_b$ & $ 0.923 \pm 0.020 $ & $ 0.934735 \pm 0.000039 $  & $ 0.934735 \pm 0.000039 $& -0.59 &  \\ 
\rowcolor[gray]{.7} & & $[ 0.934658 , 0.934812 ]$ & $[ 0.934658 , 0.934813 ]$ & &  \\ 
\rowcolor[gray]{.7} $\mathcal{A}_c$ & $ 0.670 \pm 0.027 $ & $ 0.66777 \pm 0.00021 $  & $ 0.66777 \pm 0.00022 $& 0.11 &  \\ 
\rowcolor[gray]{.7} & & $[ 0.66735 , 0.66819 ]$ & $[ 0.66735 , 0.66820 ]$ & &  \\ 
\hline 
$\mathcal{A}_s$ & $ 0.895 \pm 0.091 $ & $ 0.935642 \pm 0.000040 $  & $ 0.935642 \pm 0.000039 $& -0.43 &  \\ 
& & $[ 0.935564 , 0.935719 ]$ & $[ 0.935565 , 0.935719 ]$ & &  \\ 
BR$_{W\ell\bar\nu_\ell}$ & $ 0.10860 \pm 0.00090 $ & $ 0.108361 \pm 0.000013 $  & $ 0.108361 \pm 0.000013 $& 0.27 &  \\ 
& & $[ 0.108336 , 0.108386 ]$ & $[ 0.108336 , 0.108386 ]$ & &  \\ 
$\sin^2\theta_{\rm eff}^{ll}$ (HC) & $ 0.23150 \pm 0.00023 $ & $ 0.231523 \pm 0.000055 $  & $ 0.231523 \pm 0.000057 $& -0.13 &  \\ 
& & $[ 0.231415 , 0.231631 ]$ & $[ 0.231412 , 0.231634 ]$ & &  \\ 
\bottomrule
\end{tabular}
\end{center}
}
\caption{
  Experimental measurement, result of the global fit, individual prediction, and pull for the five input
  parameters ($\alpha_s(M_Z^2)$, $\Delta \alpha^{(5)}_{\mathrm{had}}(M_Z^2)$, $M_Z$,
  $m_t$, $m_H$), and for the set of EWPO considered in the fit, in the
  \emph{standard} scenario for $m_t$ and $m_H$. Horizontal lines
  separate groups of correlated observables. For the results of the global 
  fit and for the predictions, the $95\%$ probability range is reported in square brackets. The values in
  the column \emph{Individual Prediction} are determined without using the
  experimental information in the same row, or in the rows within the
  same block of correlated observables. Pulls are calculated
  both as individual pulls (\emph{1D Pull}) and as global pulls
  (\emph{nD Pull}) for sets of correlated observables, and are given in units of standard
  deviations. } 
\label{ewptests:tab:SMfit}
\end{table}
For each observable, Table~\ref{ewptests:tab:SMfit} gives the experimental
information used as input (\textit{Measurement}), together with the
output of the combined fit (\textit{Posterior} of the \textit{Full Fit}), the
\textit{Individual Prediction} of the same quantity, and the corresponding
1D or nD pull, as defined in
Section~\ref{ewptests:sec:ew-precision-fits-general}. 
The pulls are also summarized in Figure~\ref{ewptests:fig:pulls}.

Furthermore, as defined in
Section~\ref{ewptests:sec:ew-precision-fits-general}, 
in addition to the \emph{Individual Predictions} obtained removing
each individual observable/set of correlated observables from the fit,
one can obtain a \emph{Full Prediction} by dropping all experimental
information on EWPO and just using the information on SM
parameters. Conversely, one can obtain a \emph{Full Indirect}
determination of the SM parameters by dropping all information on all
parameters simultaneously and determining all of them from the fit to
EWPO. The results of these two extreme possibilities are reported in
Table \ref{ewptests:tab:FullIndFullPred}, together with the corresponding 1D pulls. Notice that any pull in the \emph{Full Indirect} fit purely originates from the correlation between the various EWPO induced by the SM theory, irrespective of the values of the input parameters. Therefore, any improvement of the pulls with respect to the \emph{Full Indirect} fit at fixed experimental results can only be achieved by extending the SM with additional degrees of freedom and/or interactions. It will be interesting to see how these pulls evolve when new experimental results become available in the future.
\begin{table}[!htb]
{
\begin{center}
  \scriptsize
\begin{tabular}{l|c|cc|cc}
\toprule
& Measurement & Full Indirect & Pull & Full Prediction & Pull \\
\hline 
\hline 
$\alpha_{s}(M_{Z})$ & $ 0.11873 \pm 0.00056 $ & $ 0.1196 \pm 0.0043 $ & -0.2 & $ 0.11873 \pm 0.00056 $& 0.0 \\ 
$\delta\alpha^{5}_{\mathrm{had}}$ & $ 0.02766 \pm 0.00010 $ & $ 0.02707 \pm 0.00085 $ & 0.7 & $ 0.027659 \pm 0.000099 $& 0.1 \\ 
$M_Z$ [GeV] & $ 91.1875 \pm 0.0021 $ & $ 91.208 \pm 0.040 $ & -0.5 & $ 91.1876 \pm 0.0021 $& 0.0 \\ 
$m_t$ [GeV] & $ 172.31 \pm 0.32 $ & $ 177.6 \pm 9.6 $ & -0.6 & $ 172.31 \pm 0.32 $& -0.1 \\ 
$m_H$ [GeV] & $ 125.100 \pm 0.090 $ & $ 247.1 \pm 222.8 $ & -0.2 & $ 125.099 \pm 0.090 $& -0.1 \\ 
\hline 
$M_W$ [GeV] & $ 80.3635 \pm 0.0080 $ & $ 80.3625 \pm 0.0079 $ & 0.1 & $ 80.3524 \pm 0.0055 $& 1.2 \\ 
\hline 
$\Gamma_{W}$ [GeV] & $ 2.151 \pm 0.049 $ & $ 2.0893 \pm 0.0020 $ & 1.3 & $ 2.08811 \pm 0.00048 $& 1.3 \\ 
\hline 
$\sin^2\theta_{\rm eff}^{\rm lept}(Q_{\rm FB}^{\rm had})$ & $ 0.2324 \pm 0.0012 $ & $ 0.23149 \pm 0.00013 $ & 0.8 & $ 0.231545 \pm 0.000061 $& 0.7 \\ 
\hline 
$-P_{\tau}^{\rm pol}=\mathcal{A}_\ell$ & $ 0.1465 \pm 0.0033 $ & $ 0.1473 \pm 0.0010 $ & -0.2 & $ 0.14684 \pm 0.00048 $& -0.1 \\ 
\hline 
$\Gamma_{Z}$ [GeV] & $ 2.4955 \pm 0.0023 $ & $ 2.4963 \pm 0.0018 $ & -0.3 & $ 2.49457 \pm 0.00055 $& 0.4 \\ 
$\sigma_{h}^{0}$ [nb] & $ 41.480 \pm 0.033 $ & $ 41.469 \pm 0.029 $ & 0.2 & $ 41.4875 \pm 0.0069 $& -0.2 \\ 
$R^{0}_{\ell}$ & $ 20.767 \pm 0.025 $ & $ 20.756 \pm 0.022 $ & 0.3 & $ 20.7531 \pm 0.0070 $& 0.5 \\ 
$A_{\rm FB}^{0, \ell}$ & $ 0.0171 \pm 0.0010 $ & $ 0.01627 \pm 0.00023 $ & 0.8 & $ 0.01617 \pm 0.00011 $& 0.9 \\ 
\hline 
$\mathcal{A}_{\ell}$ (SLD) & $ 0.1513 \pm 0.0021 $ & $ 0.1473 \pm 0.0010 $ & 1.7 & $ 0.14684 \pm 0.00048 $& 2.1 \\ 
$R^{0}_{b}$ & $ 0.21629 \pm 0.00066 $ & $ 0.21570 \pm 0.00036 $ & 0.8 & $ 0.21588 \pm 0.00010 $& 0.6 \\ 
$R^{0}_{c}$ & $ 0.1721 \pm 0.0030 $ & $ 0.17228 \pm 0.00017 $ & -0.1 & $ 0.172215 \pm 0.000051 $& 0.0 \\ 
$A_{\rm FB}^{0, b}$ & $ 0.0996 \pm 0.0016 $ & $ 0.10327 \pm 0.00075 $ & -2.1 & $ 0.10295 \pm 0.00034 $& -2.1 \\ 
$A_{\rm FB}^{0, c}$ & $ 0.0707 \pm 0.0035 $ & $ 0.07381 \pm 0.00058 $ & -0.9 & $ 0.07354 \pm 0.00025 $& -0.8 \\ 
$\mathcal{A}_b$ & $ 0.923 \pm 0.020 $ & $ 0.93470 \pm 0.00016 $ & -0.6 & $ 0.934724 \pm 0.000040 $& -0.6 \\ 
$\mathcal{A}_c$ & $ 0.670 \pm 0.027 $ & $ 0.66803 \pm 0.00052 $ & 0.1 & $ 0.66771 \pm 0.00022 $& 0.1 \\ 
\hline 
$\mathcal{A}_s$ & $ 0.895 \pm 0.091 $ & $ 0.935690 \pm 0.000096 $ & -0.4 & $ 0.935631 \pm 0.000040 $& -0.4 \\ 
BR$_{W\ell\bar\nu_\ell}$ & $ 0.10860 \pm 0.00090 $ & $ 0.10834 \pm 0.00010 $ & 0.3 & $ 0.108362 \pm 0.000013 $& 0.2 \\ 
$\sin^2\theta_{\rm eff}^{ll}$ (HC) & $ 0.23150 \pm 0.00023 $ & $ 0.23149 \pm 0.00013 $ & 0.0 & $ 0.231545 \pm 0.000061 $& -0.2 \\ 
\bottomrule
\end{tabular}
\end{center}
}
\caption{Results of the \emph{full indirect} determination of SM parameters using only EWPO (third column) and of the \emph{full prediction} for EWPO using only information on SM parameters (fourth column). For comparison, the input values are reported in the second column. See the text for details.}
\label{ewptests:tab:FullIndFullPred}
\end{table}
\begin{figure}[!hbt]
  \centering
  \includegraphics[width=.23\textwidth]{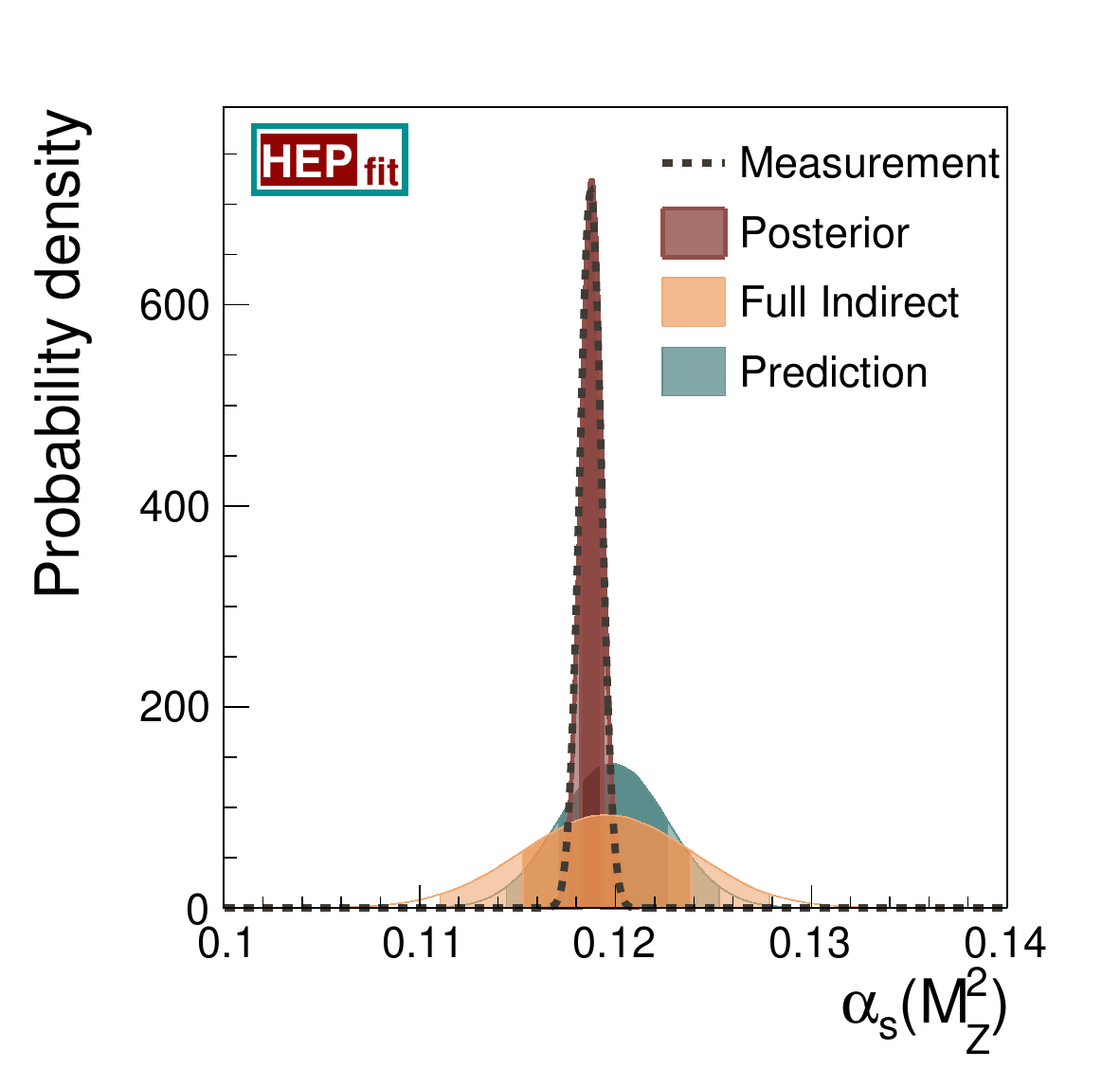}
  %\hspace{-6.3mm}
  \includegraphics[width=.23\textwidth]{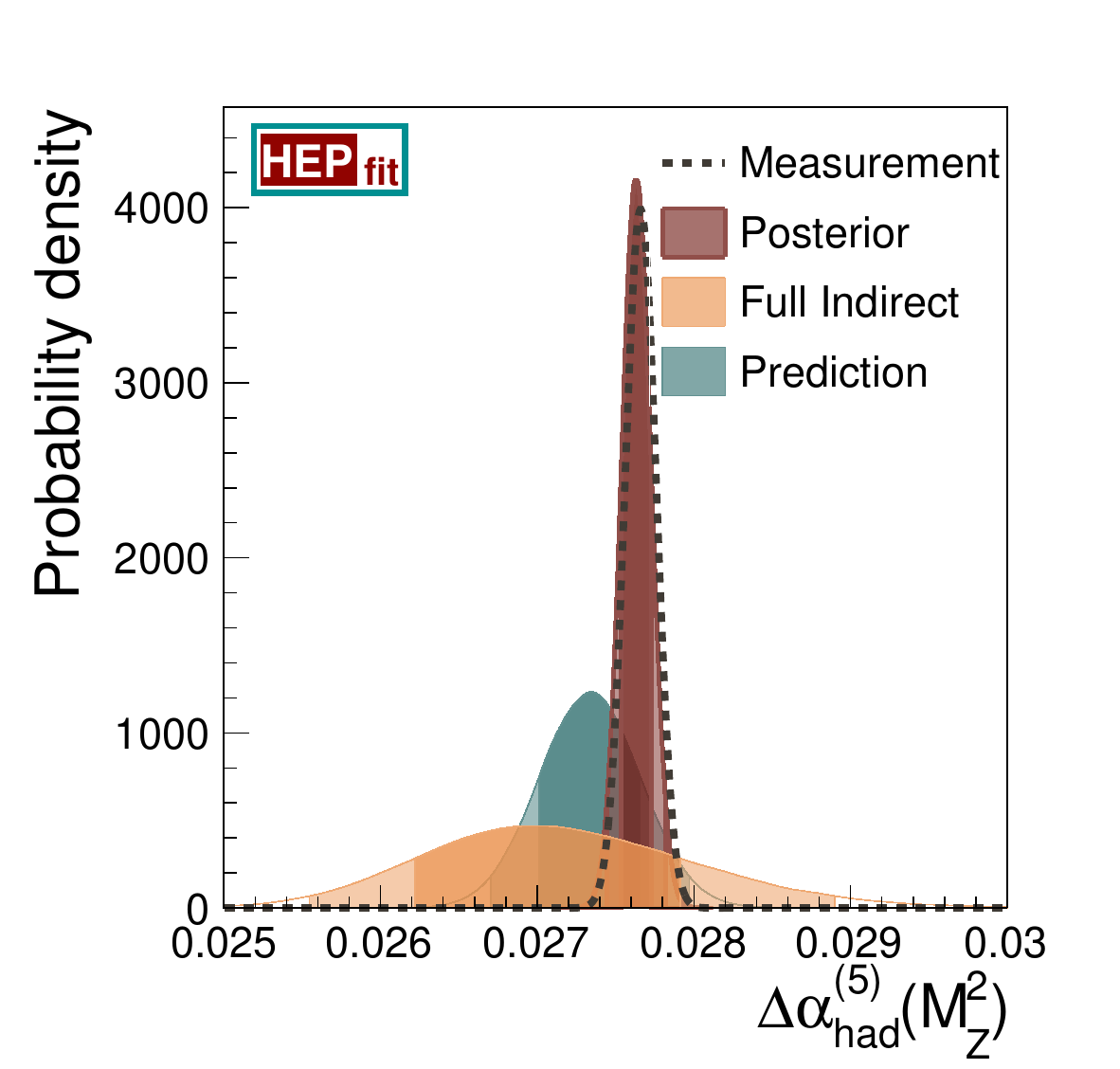}
  %\hspace{-6.3mm}
  \includegraphics[width=.23\textwidth]{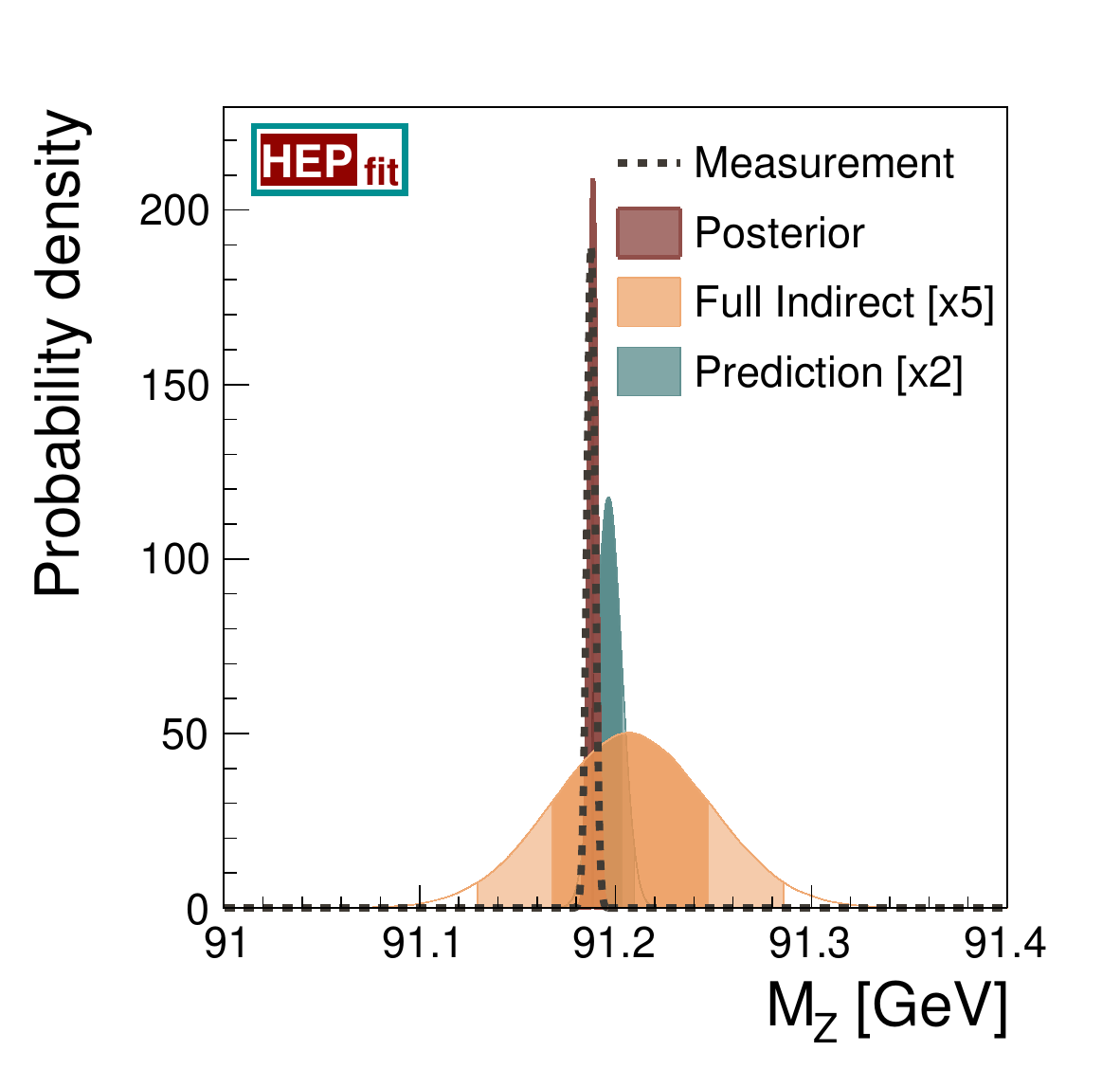}
  %\hspace{-6.3mm}
  \\
  \includegraphics[width=.23\textwidth]{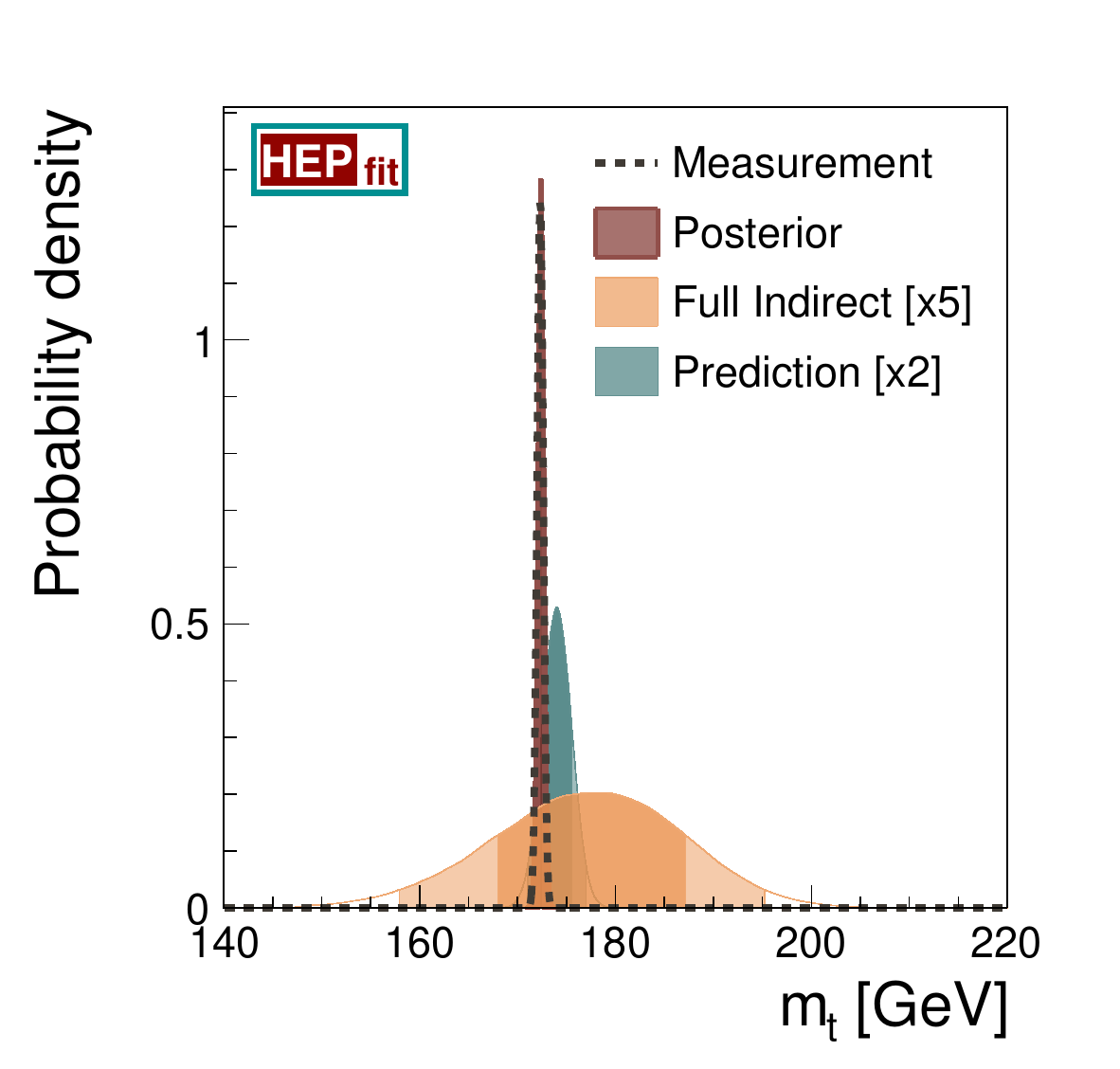}
  %\hspace{-6.3mm}
  \includegraphics[width=.23\textwidth]{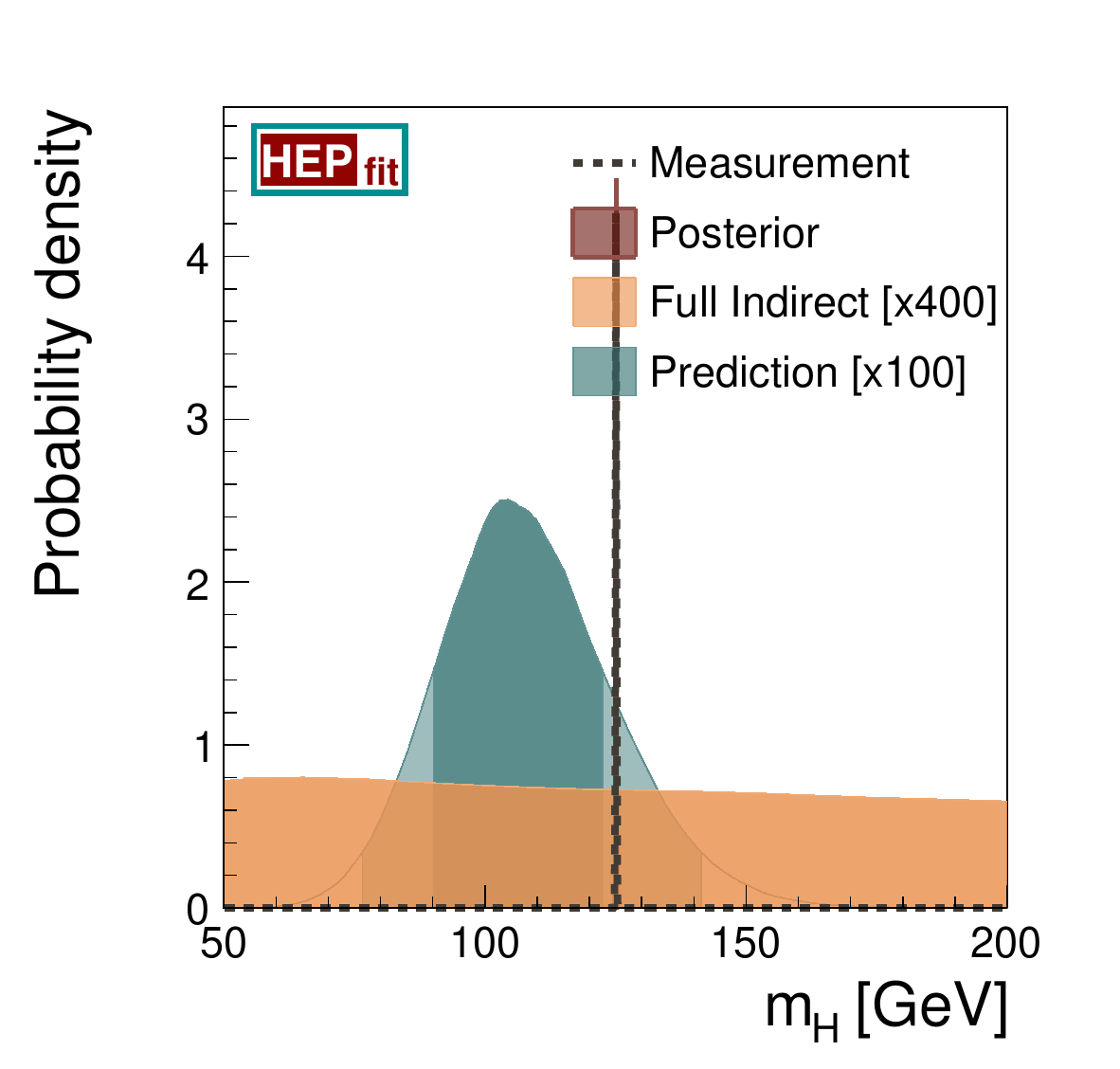}
  \caption{Comparison among the direct measurement (dashed line), the posterior of the Full Fit (red shaded area), the individual indirect determination
    (green shaded area), and the full indirect determination (orange shaded area)
    of the input parameters in the SM fit. To allow for a comparison with the other p.d.f.s, the full indirect p.d.f for the Higgs mass is truncated in the figure. Dark (light)
    regions correspond to $68\%$ ($95\%$) probability ranges.
    }
  \label{ewptests:fig:SMinputs}
\end{figure}
From the fit one also obtains posteriors for the SM parameters
$\alpha_s(M_Z^2)$, $\Delta\alpha_{\mathrm{had}}^{(5)}(M_Z^2)$, $M_Z$,
$m_t$, and $m_H$ (see Tables~\ref{ewptests:tab:SMfit} and \ref{ewptests:tab:FullIndFullPred}). They are
illustrated in Figure~\ref{ewptests:fig:SMinputs} together with their
measured values, their \textit{Indirect}, and their \textit{Full
    Indirect} determinations. We notice that while the posterior is
  always dominated by the experimental input (as desirable for fit input
  parameters), the fit would provide an indirect determination with a
  compatible result within $\sim1\sigma$, confirming the overall consistency
  of the SM.

\begin{figure}[!htb]
  \centering
  \includegraphics[width=.45\textwidth]{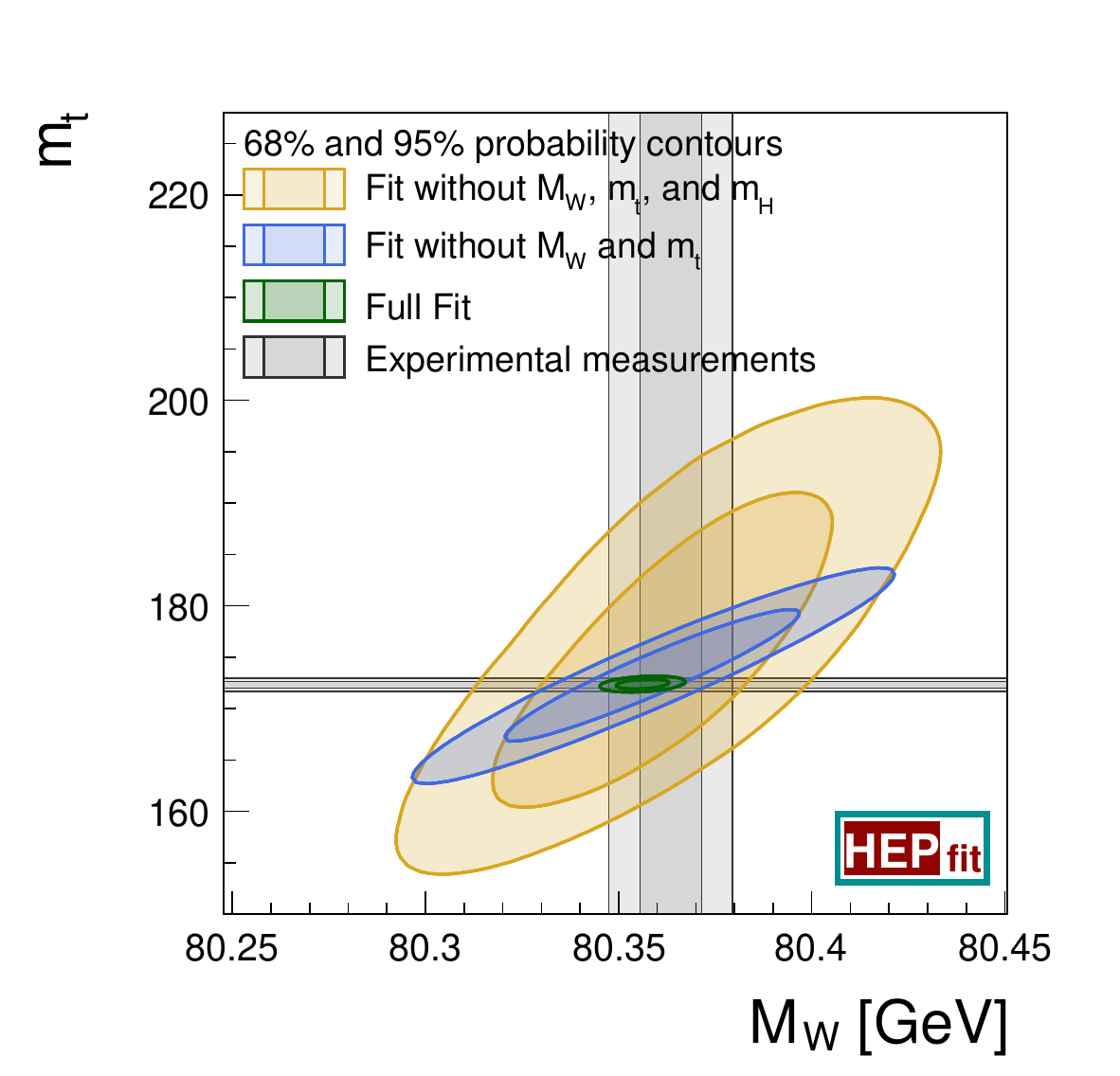}
  \hspace{-2mm}
  \includegraphics[width=.45\textwidth]{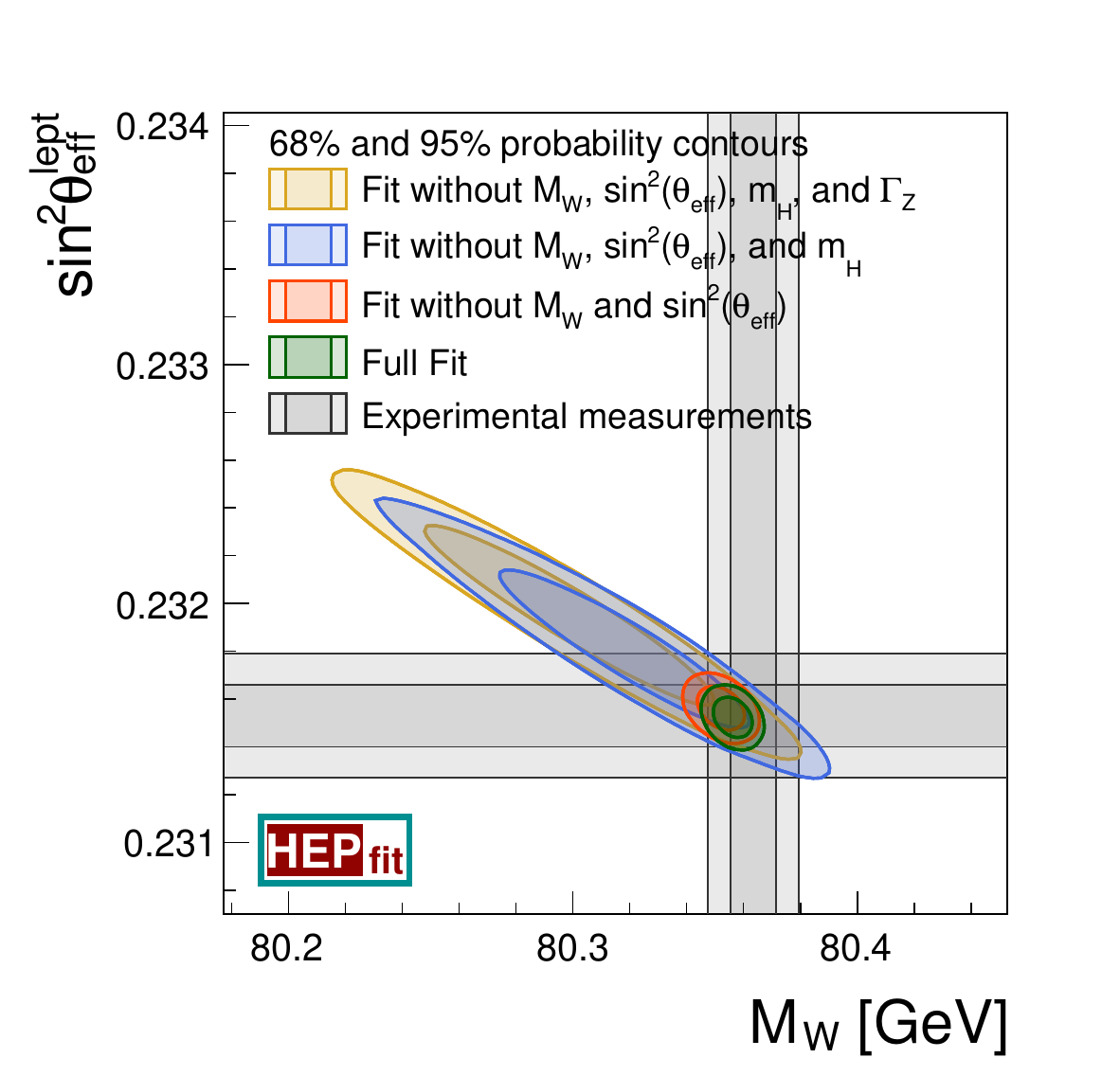}
  \caption{Impact of various constraints in the  $m_t$ vs. $M_W$ (left) and $\sin^2{\theta_{\rm eff}^{\rm lept}}$ vs. $M_W$ (right) planes. Dark (light)
    regions correspond to $68\%$ ($95\%$) probability ranges.}
  \label{ewptests:fig:2Dplots}
\end{figure}
It is instructive to investigate the impact of different experimental constraints on the fit by comparing the posterior distributions obtained omitting specific sets of experimental data. For example, as discussed in Section~\ref{ewptests:sec:sm-lagrangian}, the radiative corrections summarized in the parameter $\Delta r$  relate the $W$-boson mass to other SM parameters, with a quadratic sensitivity to the top-quark mass and a logarithmic sensitivity to the Higgs-boson mass. Thus, the experimental measurements of $m_t$ and, to a lesser extent, $m_H$ have a significant impact on the SM prediction of $M_W$. This is illustrated in the left panel of Figure~\ref{ewptests:fig:2Dplots}, where the posterior distributions in the $m_t$ vs. $M_W$ plane obtained from a global fit omitting or including the experimental information on $m_t$ and $M_W$ (or $m_t$, $M_W$, and $m_H$) are shown. It is evident from the figure that hadron collider measurements of $m_t$, $M_W$, and $m_H$ play a crucial role in testing the consistency of the SM in EW precision fits. The sensitivity of the fit to possible anomalies in the measurements of e.g.~$M_W$ has been recently demonstrated by the CDF measurement \cite{CDF:2022hxs}, which, if taken at face value and included in the pre-CMS \cite{CMS:2024lrd} average of $M_W$ measurements, would lead to a significant tension with the SM prediction of $M_W$ based on the other EWPO and SM parameters \cite{deBlas:2022hdk}. Another example of the interplay of different EWPO in constraining the SM is illustrated in the right panel of Figure~\ref{ewptests:fig:2Dplots}, where the posterior distributions in the $\sin^2\theta_{\mathrm{eff}}^\mathrm{lept}$ vs. $M_W$ plane obtained from a global fit omitting or including the experimental information for $M_W$ and $\sin^2\theta_{\mathrm{eff}}^\mathrm{lept}$ (or $M_W$, $\sin^2\theta_{\mathrm{eff}}^\mathrm{lept}$, $m_H$, and $\Gamma_Z$) are shown. To omit the experimental information on $\sin^2\theta_{\mathrm{eff}}^\mathrm{lept}$, we have removed the LEP and SLD measurements of lepton asymmetries, the tau polarization measurement at LEP and CMS, the charge asymmetry measurement at LEP, and the determinations of $\sin^2\theta_{\mathrm{eff}}^\mathrm{lept}$ from hadron collider measurements. This can be compared with the average determination of $\sin^2\theta_{\mathrm{eff}}^\mathrm{lept}$ from all these measurements, which corresponds to $\sin^2\theta_{\mathrm{eff}}^\mathrm{lept} = (0.23152 \pm 0.00013)$ and is reported in Figure~\ref{ewptests:fig:2Dplots} as a gray band. Differently from the case of $M_W$, radiative corrections to $\sin^2\theta_{\mathrm{eff}}^\mathrm{lept}$ are not quadratically sensitive to $m_t$, so that the logarithmic sensitivity to $m_H$ plays a more important role in constraining the SM prediction of $\sin^2\theta_{\mathrm{eff}}^\mathrm{lept}$. This is evident from comparing the posteriors obtained omitting or including the experimental information on $m_H$ in the fit.

\section{The EW precision fit beyond the SM: S,T,U and SMEFT}
\label{ewptests:sec:ew-precision-fits-bsm}

To illustrate the impact of EWPO in constraining physics beyond the SM we consider two widely adopted frameworks: the case of \textit{oblique} models in which NP effects mainly appear in the gauge-boson vacuum-polarization corrections, also called \textit{oblique}\footnote{This term was introduced to
  distinguish the vacuum-polarization-corrections from the
  \textit{direct} vertex and box corrections
which modify the form of the interactions themselves.} corrections,  and the case of the \textit{Standard Model Effective Field Theory} (SMEFT) where the effect of non-SM interactions induced by NP living at a scale $\Lambda$ much higher than the EW scale is described through a systematic expansion in higher-dimension operators that extend the SM Lagrangian.

\subsection{The case of Oblique Models: S,T,U parametrization}
\label{ewptests:sec:bsm-stu}
In several models beyond the SM, the dominant NP effects appear in the
corrections to the EW gauge-boson vacuum polarization~\cite{Kennedy:1988sn,Kennedy:1988rt}, i.e. in
oblique corrections. If the mass
scale of NP is sufficiently higher than the EW scale, these oblique
corrections are effectively described by the three independent
parameters $S$, $T$ and $U$~\cite{Peskin:1990zt,Peskin:1991sw}:
\begin{eqnarray}
S &=& 
-16\pi \Pi^{\mathrm{NP}\prime}_{30}(0)
= 16\pi
\left[\Pi^{\mathrm{NP}\prime}_{33}(0) - \Pi^{\mathrm{NP}\prime}_{3Q}(0)
\right],
\\
T &=& \frac{4\pi}{s_W^2c_W^2 M_Z^2}
\left[\Pi^{\mathrm{NP}}_{11}(0) - \Pi^{\mathrm{NP}}_{33}(0)
\right],
\\
U &=& 16\pi
\left[\Pi^{\mathrm{NP}\prime}_{11}(0) - \Pi^{\mathrm{NP}\prime}_{33}(0)
\right],
\end{eqnarray}
where $\Pi^{\rm NP}_{XY}$ with $X,Y=0,1,3,Q$ denotes new-physics
contribution to the vacuum-polarization amplitude of the EW gauge bosons
defined, e.g., in Ref.~\cite{Peskin:1991sw}, 
$\Pi'_{XY}(q^2) = d\Pi_{XY}(q^2)/dq^2$, while all other parameters
are SM parameters introduced in Section~\ref{ewptests:sec:sm-lagrangian}.
NP contributions to a given EWPO $\mathcal{O}$ can then be
expressed in terms of oblique parameters and, at the linear order in
$S$, $T$, and $U$, can be written as
\begin{equation}
\mathcal{O} = \mathcal{O}_{\rm SM} + \mathcal{O}_{NP}(S,T,U)\,,
\end{equation} 
where $S=T=U=0$ in the SM
~\cite{Peskin:1990zt,Peskin:1991sw,Maksymyk:1993zm,Burgess:1993mg,Burgess:1993vc}. 

More specifically, all EWPO can be expressed in
terms of the following combinations of oblique parameters: 
\begin{eqnarray}
  \label{eq:abc}
  A &=& S - 2c_W^2\, T - \frac{(c_W^2-s_W^2)\,U}{2s_W^2}\,,
  \nonumber \\
  B &=& S - 4c_W^2 s_W^2\, T\,,\\
  C &=& -10(3-8s_W^2)\,S + (63-126s_W^2-40s_W^4)\,T\,,\nonumber
\end{eqnarray}
where it is interesting to notice that the parameter $C$ describes the new-physics contribution to
$\Gamma_{Z}$, the parameter $A$ (the only one containing $U$)
describes the NP contribution to $M_W$ and $\Gamma_W$, and NP
contributions to all other EWPO are proportional to $B$. By fitting the three oblique parameters together
with the SM parameters to EW precision data one can constrain the
allowed values of $S$, $T$, and $U$. 
The fit results are summarized in table~\ref{ewptests:tab:STU}, which displays the high degree of consistency of the EW precision data with the SM, with oblique parameters constrained in the few percent range.

\begin{figure}[t]
  \centering
  \includegraphics[width=.24\textwidth]{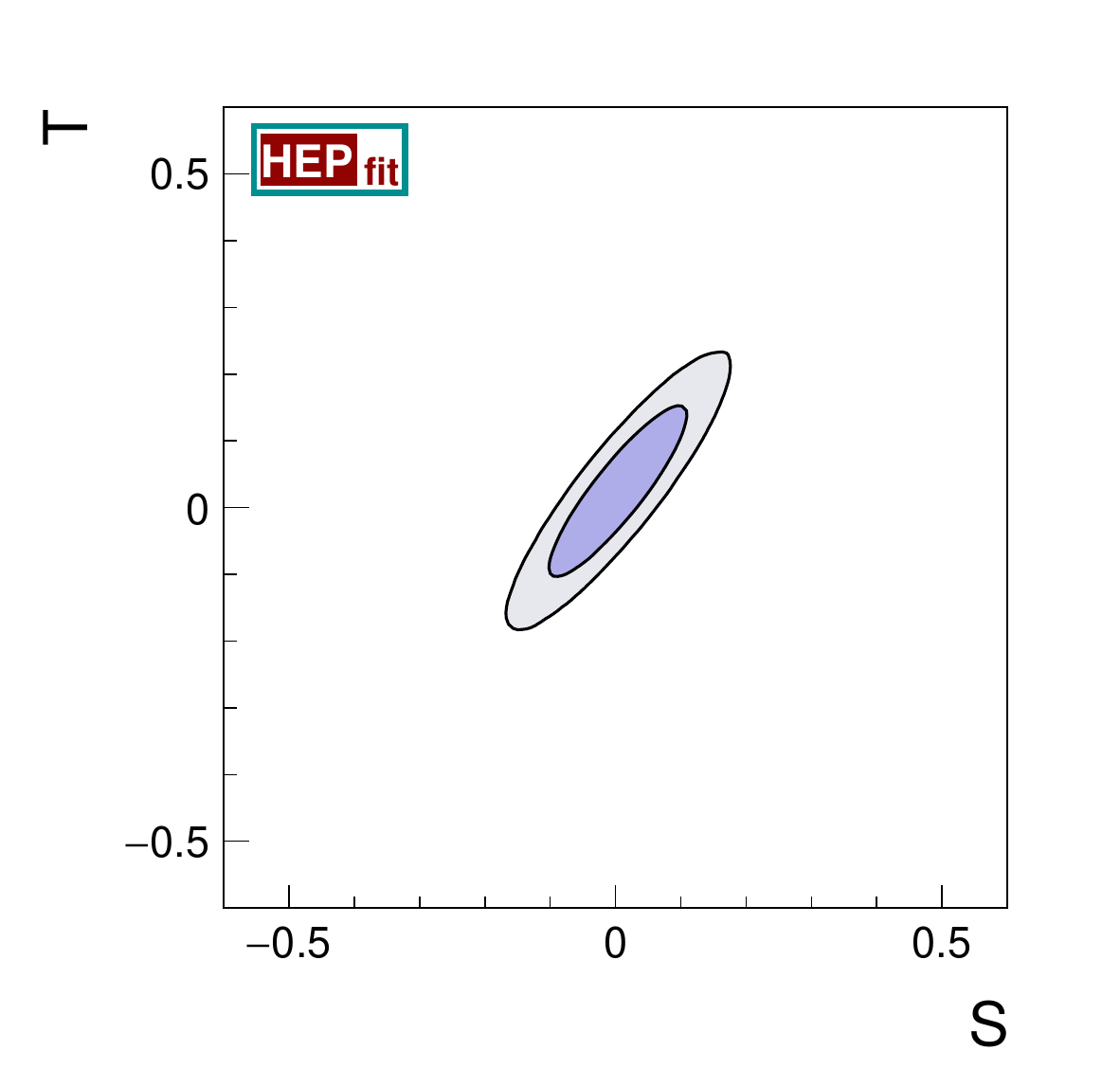}
  \includegraphics[width=.24\textwidth]{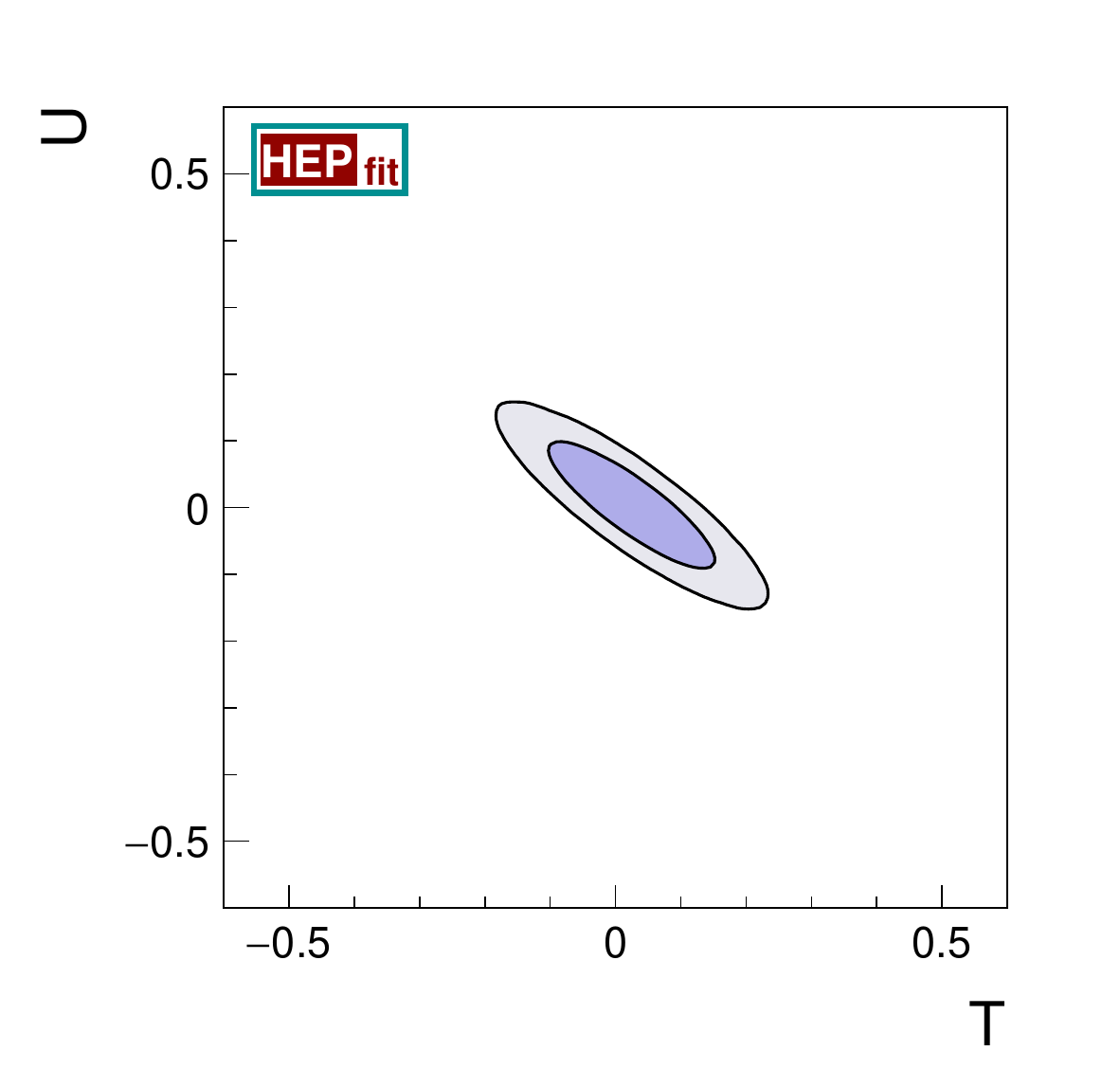}
  \includegraphics[width=.24\textwidth]{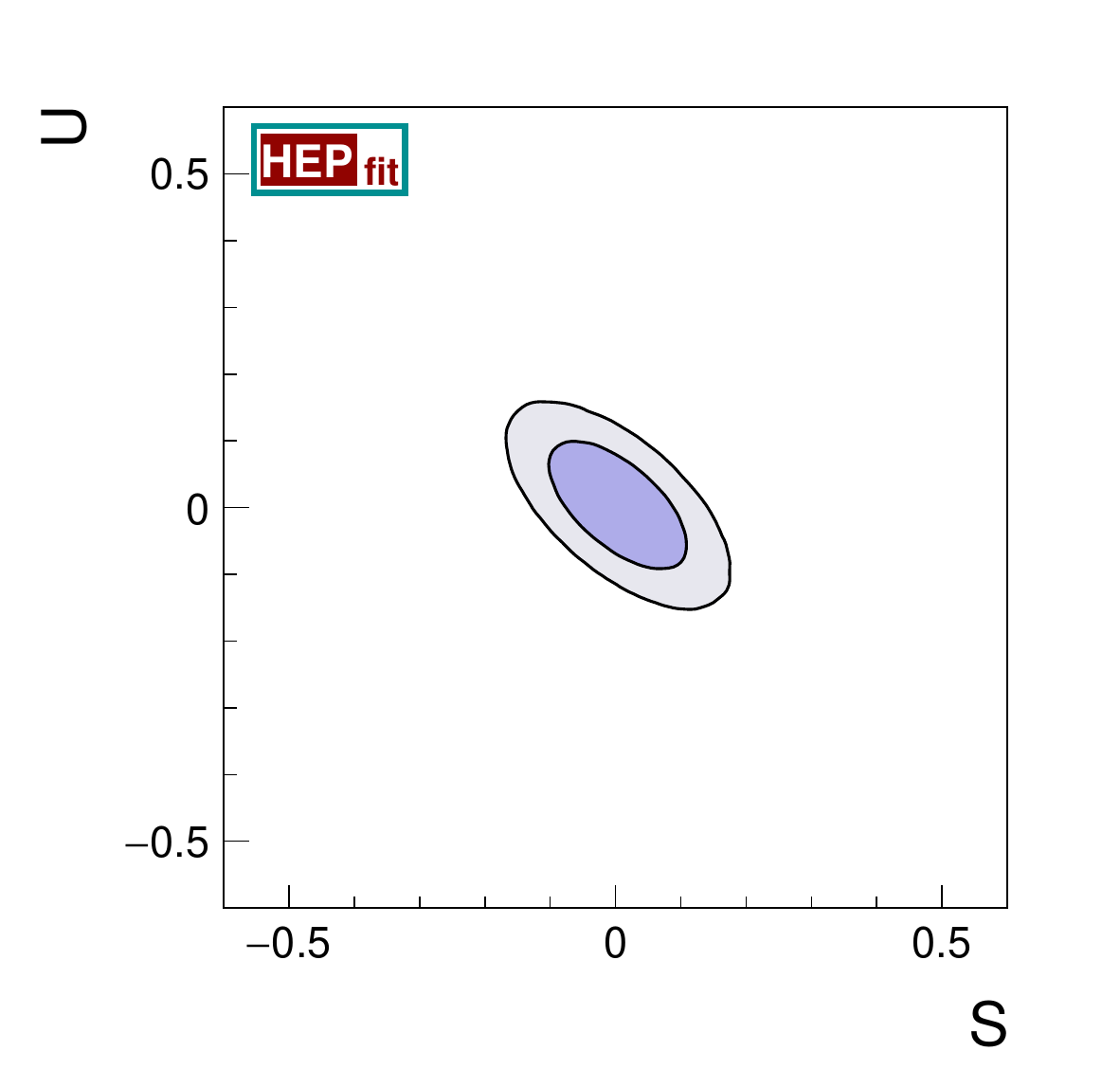}
  \includegraphics[width=.24\textwidth]{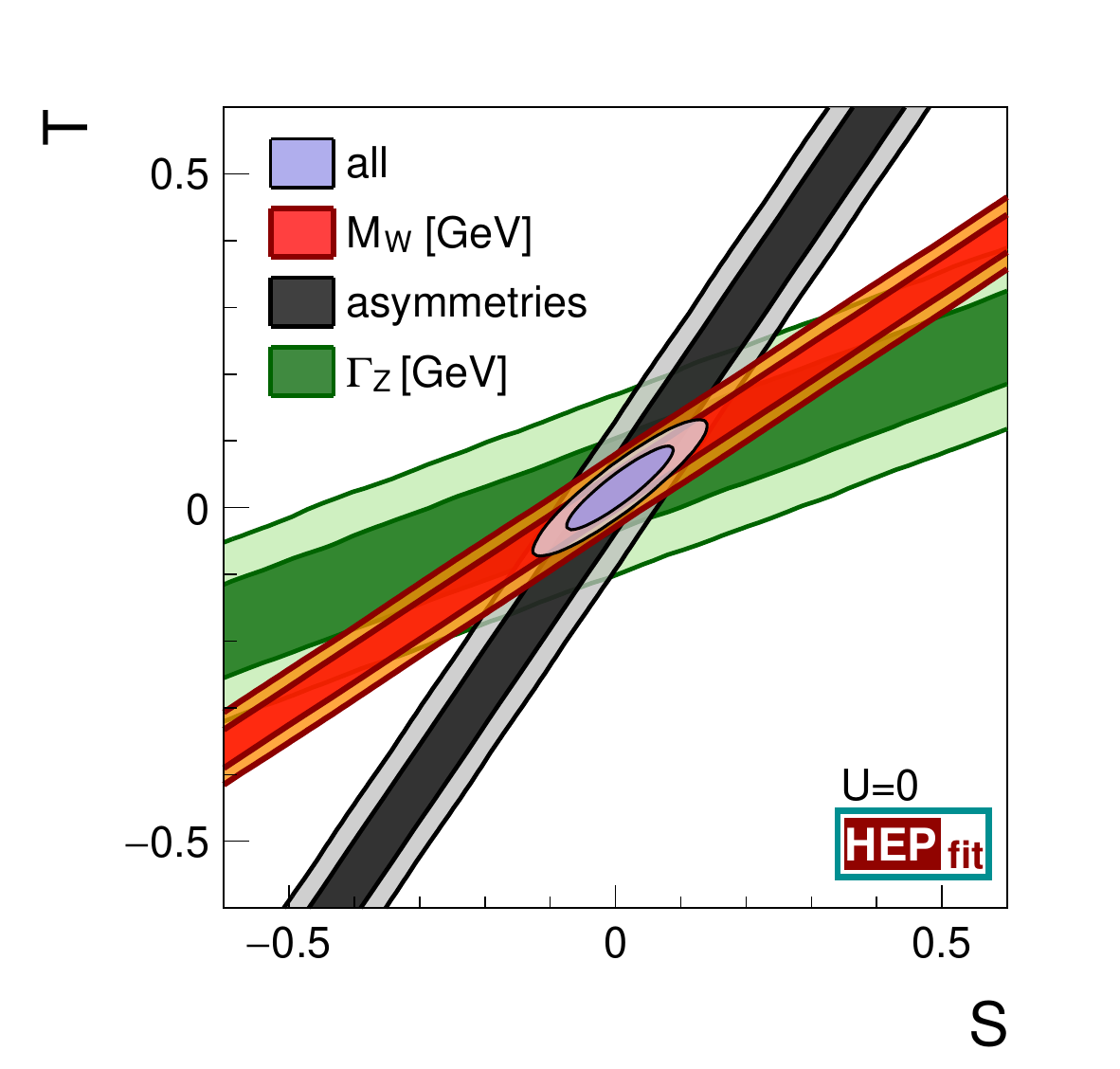}
   \caption{Two-dimensional posteriors for oblique parameters from the EW precision fit. Right panel: scenario with $U=0$. Other panels: scenarios with $U$ free. Dark (light)
    regions correspond to $68\%$ ($95\%$) probability ranges. For the $U=0$ case, the colored bands represent the constraints from $M_W$ and $\Gamma_W$ alone (red), from $\Gamma_Z$ alone (green), and from all other EWPO (gray).}
  \label{ewptests:fig:STU}
\end{figure}

\begin{table}[tp]
\centering
\begin{tabular}{c|cccr|ccc} 
\hline
 & 
\multicolumn{4}{c|}{\textit{STU} fit} & \multicolumn{3}{c}{\textit{ST} fit with $U=0$}  
\\
\hline
Parameter & Value & \multicolumn{3}{c|}{Correlation} & Value & \multicolumn{2}{c}{Correlation} \\ 
\hline
$S$ &
$0.007\pm 0.061$ & $1.00$ & $0.90$ & $-0.63$ &$0.007\pm 0.055$ & $ 1.00$ &  $0.92$ 
\\
$T$ &
$0.027\pm 0.088$ & & $1.00$ & $-0.87$ & $0.028\pm 0.042$ & & $1.00$
\\
$U$ &
$0.005\pm 0.064$ & & & $1.00$ & --- & &
\\
\hline
\end{tabular}
\caption{Fit results for the oblique parameters with floating $U$ or
  fixing $U=0$, using the large-$m_t$ expansion or with the results of 
  Refs.~\cite{Freitas:2012sy,freitasprivate} for the two-loop
  fermionic EW corrections to $\rho_Z^f$. In the latter case, we do
  not consider constraints from $\Gamma_Z$, $\sigma^0_h$ and $R_\ell^0$.
}
\label{ewptests:tab:STU}
\end{table}
The two-dimensional probability
distribution for different pairs of oblique parameters are shown in
the first three  plots of figure~\ref{ewptests:fig:STU}. If one fixes $U=0$, which is the case in many NP models where $U \ll
S,T$, the fit yields the results reported in the right part of
table~\ref{ewptests:tab:STU}. The corresponding
two-dimensional distribution is given in the right plot in
figure~\ref{ewptests:fig:STU}.  

Before the discovery of the Higgs boson, and the precise measurement of the top-quark mass, a model-independent parameterization of SM radiative corrections plus NP effects in EW gauge boson self-energies was suggested in refs.~\cite{Altarelli:1990zd,Altarelli:1991fk}, introducing the $\epsilon_i$ ($i=1,2,3$) parameters, and later extended to also include $Zb \bar b$ vertex corrections with the introduction of the $\epsilon_b$ parameter. After the Higgs-boson discovery, and the precise measurement of $m_t$, there is no practical advantage in using the $\epsilon_i$ parameters instead of $S$, $T$, and $U$, and a generalization to include non-oblique corrections as well as new interactions is achieved in general effective field theory frameworks such as the SMEFT one discussed in the next section. 

\subsection{The case of the SMEFT: going beyond EW precision fits}
\label{ewptests:sec:bsm-smeft}

\begin{table}[!b] 
  \centering
  \renewcommand{\arraystretch}{1.5}
  \scalebox{0.87}{
  \begin{tabular}{|c|c|c|c|} 
  \hline
  $X^3$ & 
  $\phi^6$~ and~ $\phi^4 D^2$ &
  $\psi^2\phi^3$ & $X^2 \phi^2$ \\
  \hline
  $\mathcal{O}_G=f^{ABC} G_\mu^{A\nu} G_\nu^{B\rho} G_\rho^{C\mu} $ &  
  $\mathcal{O}_\phi=(\phi^\dag \phi)^3$ &
  $\mathcal{O}_{e\phi}^{[pr]}=(\phi^\dag \phi)(\bar l_p \phi e_r )$ & 
    $\mathcal{O}_{\phi G}=\phi^\dag \phi\, G^A_{\mu\nu} G^{A\mu\nu}$ \\
  $\mathcal{O}_{W}=\varepsilon^{IJK} W_\mu^{I\nu} W_\nu^{J\rho} W_\rho^{K\mu}$ &   
  $\mathcal{O}_{\phi\Box}=(\phi^\dag \phi) \Box(\phi^\dag \phi)$ &
  $\mathcal{O}_{u\phi}^{[pr]}=(\phi^\dag \phi)(\bar q_p \widetilde{\phi} u_r )$ & 
    $\mathcal{O}_{\phi W}=\phi^\dag \phi\, W^I_{\mu\nu} W^{I\mu\nu}$
\\  
  &
  $\mathcal{O}_{\phi D}=\left(\phi^\dag D^\mu \phi\right)^\star \left(\phi^\dag D_\mu \phi\right)$ &
  $\mathcal{O}_{d\phi}^{[pr]}=(\phi^\dag \phi)(\bar q_p \phi d_r )$ &
   $\mathcal{O}_{\phi B}=\phi^\dag \phi\, B_{\mu\nu} B^{\mu\nu}$ \\ 
  & & & $\mathcal{O}_{\phi WB}= \phi^\dag \tau^I \phi\, W^I_{\mu\nu} B^{\mu\nu}$ \\
  \hline
 $\psi^2 X \phi$ &
 $\psi^2 \phi^2 D$& 
  $(\bar RR)(\bar RR)$ &
  $(\bar LL)(\bar RR)$\\ 
  \hline
  $\mathcal{O}_{e W}^{[pr]}=(\bar l_p \sigma^{\mu\nu} e_r) \tau^I \phi W_{\mu\nu}^I$ &
  $\mathcal{O}_{\phi l}^{(1)[pr]}=(\phi^\dag i\!\!\stackrel{\leftrightarrow}{D}_\mu \phi) (\bar l_p \gamma^\mu l_r)$&
  $\mathcal{O}_{ee}^{[prst]}=(\bar e_p \gamma_\mu e_r)(\bar e_s \gamma^\mu e_t)$ &
  $\mathcal{O}_{le}^{[prst]}=(\bar l_p \gamma_\mu l_r)(\bar e_s \gamma^\mu e_t)$ \\
  $\mathcal{O}_{e B}^{[pr]}=(\bar l_p \sigma^{\mu\nu} e_r) \phi B_{\mu\nu}$ &
  $\mathcal{O}_{\phi l}^{(3)[pr]}=(\phi^\dag i\!\! \stackrel{\leftrightarrow}{D^I_\mu} \phi) (\bar l_p \tau^I \gamma^\mu l_r)$&
  $\mathcal{O}_{uu}^{[prst]}=(\bar u_p \gamma_\mu u_r)(\bar u_s \gamma^\mu u_t)$ &
  $\mathcal{O}_{lu}^{[prst]}=(\bar l_p \gamma_\mu l_r)(\bar u_s \gamma^\mu u_t)$\\
  $\mathcal{O}_{uG}^{[pr]}=(\bar q_p \sigma^{\mu\nu} T^A u_r) \widetilde{\phi}\, G_{\mu\nu}^A$ &
  $\mathcal{O}_{\phi e}^{[pr]}= (\phi^\dag i\!\!\stackrel{\leftrightarrow}{D}_\mu \phi) (\bar e_p \gamma^\mu e_r)$&
  $\mathcal{O}_{dd}^{[prst]}=(\bar d_p \gamma_\mu d_r)(\bar d_s \gamma^\mu d_t)$ &
  $\mathcal{O}_{ld}^{[prst]}=(\bar l_p \gamma_\mu l_r)(\bar d_s \gamma^\mu d_t)$\\  
  $\mathcal{O}_{uW}^{[pr]}=(\bar q_p \sigma^{\mu\nu} u_r) \tau^I \widetilde{\phi}\, W_{\mu\nu}^I$ &
  $\mathcal{O}_{\phi q}^{(1)[pr]}=(\phi^\dag i\!\!\stackrel{\leftrightarrow}{D}_\mu \phi) (\bar q_p \gamma^\mu q_r)$&
  $\mathcal{O}_{eu}^{[prst]}=(\bar e_p \gamma_\mu e_r)(\bar u_s \gamma^\mu u_t)$ &
  $\mathcal{O}_{qe}^{[prst]}=(\bar q_p \gamma_\mu q_r)(\bar e_s \gamma^\mu e_t)$ \\
  $\mathcal{O}_{u B}^{[pr]}=(\bar q_p \sigma^{\mu\nu} u_r) \widetilde{\phi}\, B_{\mu\nu}$&
  $\mathcal{O}_{\phi q}^{(3)[pr]}=(\phi^\dag i \!\!\stackrel{\leftrightarrow}{D^I_\mu} \phi) (\bar q_p \tau^I \gamma^\mu q_r)$&
  $\mathcal{O}_{ed}^{[prst]}=(\bar e_p \gamma_\mu e_r)(\bar d_s\gamma^\mu d_t)$ &
  $\mathcal{O}_{qu}^{(1)[prst]}=(\bar q_p \gamma_\mu q_r)(\bar u_s \gamma^\mu u_t)$ \\
  $\mathcal{O}_{dG}^{[pr]}=(\bar q_p \sigma^{\mu\nu} T^A d_r) \phi\, G_{\mu\nu}^A$ & 
  $\mathcal{O}_{\phi u}^{[pr]}=(\phi^\dag i\!\!\stackrel{\leftrightarrow}{D}_\mu \phi) (\bar u_p \gamma^\mu u_r)$&
  $\mathcal{O}_{ud}^{(1)[prst]}=(\bar u_p \gamma_\mu u_r)(\bar d_s \gamma^\mu d_t)$ &
  $\mathcal{O}_{qu}^{(8)[prst]}=(\bar q_p \gamma_\mu T^A q_r)(\bar u_s \gamma^\mu T^A u_t)$ \\ 
  $\mathcal{O}_{dW}^{[pr]}=(\bar q_p \sigma^{\mu\nu} d_r) \tau^I \phi\, W_{\mu\nu}^I$ &
  $\mathcal{O}_{\phi d}^{[pr]}=(\phi^\dag i\!\!\stackrel{\leftrightarrow}{D}_\mu \phi) (\bar d_p \gamma^\mu d_r)$& 
  $\mathcal{O}_{ud}^{(8)[prst]}=(\bar u_p \gamma_\mu T^A u_r)(\bar d_s \gamma^\mu T^A d_t)$ &
  $\mathcal{O}_{qd}^{(1)[prst]}=(\bar q_p \gamma_\mu q_r)(\bar d_s \gamma^\mu d_t)$\\
  $\mathcal{O}_{dB}^{[pr]}=(\bar q_p \sigma^{\mu\nu} d_r) \phi\, B_{\mu\nu}$ &
  $\mathcal{O}_{\phi ud}^{[pr]}=(\widetilde{\phi}^\dag iD_\mu \phi)(\bar u_p \gamma^\mu d_r)$ &&
  $\mathcal{O}_{qd}^{(8)[prst]}=(\bar q_p \gamma_\mu T^A q_r)(\bar d_s \gamma^\mu T^A d_t)$\\  
  \hline 
  $(\bar LL)(\bar LL)$ & $(\bar LR)(\bar LR)$ & 
  $(\bar LR)(\bar RL)$ &
  \\
  \hline
  $\mathcal{O}_{ll}^{[prst]}=(\bar l_p \gamma_\mu l_r)(\bar l_s \gamma^\mu l_t)$ &
  $\mathcal{O}_{quqd}^{(1)[prst]}=(\bar q_p^i u_r)\epsilon_{ij}(\bar q_s^j d_t)$ &
  $\mathcal{O}_{ledq}^{[prst]}=(\bar l_p^i  e_r)(\bar d_s q_{ti})$ &
  \\
  $\mathcal{O}_{qq}^{(1)[prst]}=(\bar q_p \gamma_\mu q_r)(\bar q_s \gamma^\mu q_t)$ &
  $\mathcal{O}_{quqd}^{(8)[prst]}=(\bar q_p^i T^A u_r)\epsilon_{ij}(\bar q_s^j T^A d_t)$&
   &
  \\
  $\mathcal{O}_{qq}^{(3)[prst]}=(\bar q_p \gamma_\mu \tau^I q_r)(\bar q_s \gamma^\mu \tau^I q_t)$  &
  $\mathcal{O}_{lequ}^{(1)[prst]}=(\bar l_p^i e_r)\epsilon_{ij}(\bar q_s^j u_t)$ &
 &
  \\
  $\mathcal{O}_{lq}^{(1)[prst]}=(\bar l_p \gamma_\mu l_r)(\bar q_s \gamma^\mu q_t)$ &
  $\mathcal{O}_{lequ}^{(3)[prst]}=(\bar l_p^i \sigma_{\mu\nu} e_r)\epsilon_{ij}(\bar q_s^j \sigma^{\mu\nu} u_t)$&
   &
  \\
  $\mathcal{O}_{lq}^{(3)[prst]}=(\bar l_p \gamma_\mu \tau^I l_r)(\bar q_s \gamma^\mu \tau^I q_t)$ &&&\\ 
  \hline
  \end{tabular}

  }
  \caption{Dimension-6 operators in the Warsaw basis, adapted from ref.~\cite{Grzadkowski:2010es}. We collectively denote by $X$ the field
    strength tensors of the SM gauge group ($X=G^{\mu\nu},W^{\mu\nu},B^{\mu\nu}$), by $\psi$ a generic
    fermion field, and by $\phi$ the scalar field of the
    SM. Among the fermion fields, $q$ and
    $l$ denote $\mathrm{SU}(2)_L$ left-handed doublets while $u,d,e$ $\mathrm{SU}(2)_L$ right-handed
    singlets, with family indices specified by the superscript in square
    brackets and chirality labels suppressed for conciseness.\label{ewptests:tab:SMEFTops}}
\end{table}

With the LHC entering the high-luminosity phase, and thanks to the experimental progress that has allowed to achieve for many observables an experimental precision much better than initially expected, several high-$p_T$ measurements of Higgs-boson, top-quark and Drell-Yan observables are now reaching a sensitivity to NP effects comparable to that of EWPO. Precision tests of the SM can therefore be extended beyond the traditional EW precision fit to include a larger set of observables, and a more general class of NP effects. A systematic and model-independent framework to describe such effects is provided by the SMEFT, where the SM Lagrangian is extended with a complete set of higher-dimension operators constructed from SM fields and respecting the SM gauge symmetries~\cite{Buchmuller:1985jz,Grzadkowski:2010es}:
\begin{equation}
\label{eq:smeft-lagrangian}
\mathcal{L}_{\mathrm{SMEFT}}=\mathcal{L}_{\mathrm{SM}}+\sum_{d>4}\frac{1}{\Lambda^{d-4}}\sum_{i}C_i^{(d)}\mathcal{O}_i^{(d)},
\end{equation}
where ${\cal O}_i^{(d)}$ are Lorentz and (SM)-gauge invariant operators of canonical mass dimension $d$ built only of SM fields, $i$ collectively labels each operator according to its
field composition and quantum numbers, and $\Lambda$ is the cutoff of the effective theory indicatively identified with the scale of NP. Considering only observables involving energies $E \ll \Lambda$, we can safely truncate the expansion in eq.~\eqref{eq:smeft-lagrangian} to the lowest-dimension operators beyond the SM ones, i.e.~to $d=6$ (at $d=5$ there is only one operator~\cite{Weinberg:1979sa}, which gives rise to Majorana neutrino masses and is not relevant for lepton-number conserving processes). For the same reason, the leading NP effects in physical observables can be consistently computed at linear order in the Wilson coefficients $C_i^{(6)}$ of the dimension-six operators.

As discussed in detail in ref.~\cite{Grzadkowski:2010es}, for a single generation of fermions there are 59 independent baryon-number conserving $d=6$ operators (see Table \ref{ewptests:tab:SMEFTops}, where operators involving dual gauge boson field strength tensors have been omitted). Taking into account three generations of fermions, the total number of independent hermitian operators increases to 2499. Among them, however, one finds operators mediating flavor-changing neutral-current processes, which are severely constrained by low-energy flavor-physics observables. Therefore, to allow for a relatively low NP scale $\Lambda$ without violating existing experimental constraints, one has to assume a specific flavor structure for the Wilson coefficients at the scale $\Lambda$. The number of independent operators depends on the specific flavor assumptions adopted. The assumption with minimal number of independent coefficients is that of the full $U(3)^5$ flavor symmetry of SM gauge interactions, which, assuming also no new sources of CP violation, reduces the number of independent hermitian operators to $41$. Relaxing the flavor symmetry to $U(2)^5$, i.e.~treating the third generation of fermions differently from the first two ones, increases the number of independent hermitian operators to $124$ (still assuming CP-conserving NP). 

\begin{figure}[!tb]
    \centering
    \includegraphics[width=1.\linewidth]{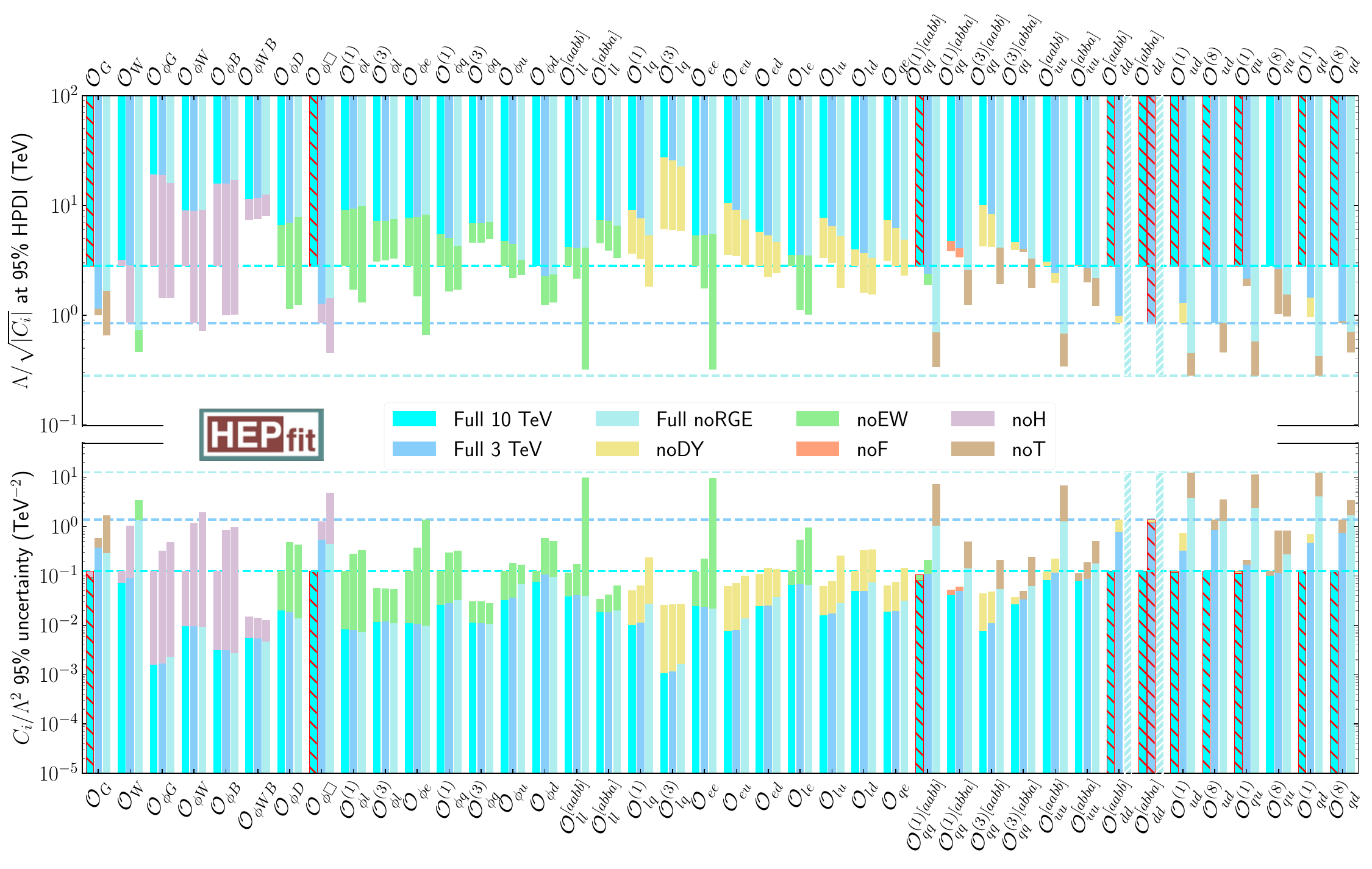}
    \caption{Results from individual fits in the $U(3)^5$ flavor symmetric SMEFT. For each coefficient $C_i$, the top panel shows the scale of NP allowed by the data at 95\% probability (normalized by the square root of the maximum of the 95\% probability interval for $|C_i|$). The bottom panel shows the width of the 95\% probability range divided by two. Both panels show results for the three cases 1) $\Lambda=10$~TeV with RG evolution, 2) $\Lambda=3$~TeV with RG evolution, and 3) $\Lambda=1$~TeV with no RG evolution. Furthermore, in each case we also include results obtained removing the most constraining data set for that particular coefficient. The color code is as explained in the legend. The horizontal lines indicate the maximum value allowed for each Wilson coefficient in the fit, corresponding to the perturbativity limit $4\pi$, for the different values chosen for the NP scale $\Lambda$. The cases in which the $95\%$ probability interval touches the prior's edges are hatched with red diagonal lines. When the posterior distribution of a coefficient is completely flat the $95\%$ probability interval is hatched with diagonal white lines. Taken from ref.~\cite{deBlas:2025xhe}.
    }
    \label{ewptests:fig:full_fit_ind_U3}
\end{figure}

Among the $41$ independent coefficients in the $U(3)^5$ flavor-symmetric scenario, when including renormalization group (RG) evolution at leading order from the scale $\Lambda$ down to the EW scale, one finds that, while about one third of the operators get the strongest constraint from EWPO, the other two thirds are mainly constrained by Drell-Yan, Higgs-boson, and top-quark observables, with comparable sensitivity to the NP scale $\Lambda$. This is illustrated in Figure~\ref{ewptests:fig:full_fit_ind_U3}, where the results of individual fits for each Wilson coefficient are shown for three different choices of the NP scale $\Lambda$.\footnote{At linear order with tree-level matrix elements the coefficient of $O_\phi$ cannot be constrained, therefore it is not shown in Figure ~\ref{ewptests:fig:full_fit_ind_U3}.} The impact of RG effects is clearly visible when comparing the results for $\Lambda=1$~TeV (no RG evolution included) with those for $\Lambda=3$~TeV and $\Lambda=10$~TeV (including RG evolution). It is evident from Figure \ref{ewptests:fig:full_fit_ind_U3} that the future of precision tests of the SM lies in the combination of EWPO with the much broader range of precision measurements that will become available with increasing accuracy at the HL-LHC and future colliders. 

The synergy between EWPO and other precision measurements is even more pronounced in the $U(2)^5$ flavor-symmetry scenario, where flavor observbles also play a crucial role in constraining several Wilson coefficients. We refer the reader to the recent analyses presented in refs.~\cite{Allwicher:2023shc,Bartocci:2024fmm,deBlas:2025xhe} for further details.

%%%%%%%%%%%%%%%%%%%%%%%%%%%%%%%%%%%%%%%%%
%% Mandatory: A concluding paragraph summing up your main points in the chapter
%% Optional: Also include big questions in the field that are still to be answered. What topics/methods/questions are researchers like to focus on next?
\section{Conclusions}
\label{ewptests:sec:conclusions}
%Mandatory: A concluding paragraph summing up the main points of your chapter.\\
%Optional: Also include big questions in the field that are still be be answered. 
%What topics/methods/questions are our colleagues like to focus on next?

In this chapter we have reviewed the current status of EW precision tests of the SM and their impact in constraining NP effects. The impressive experimental progress achieved in the measurements of EWPO at LEP, SLD, Tevatron, and LHC, together with the theoretical effort in computing higher-order corrections to EWPO within the SM, have allowed to perform very precise tests of the SM at the quantum level. The results of the global EW precision fit show a remarkable consistency of the SM predictions with experimental data, confirming the SM as the correct theory of EW interactions up to the energy scales currently probed by collider experiments. The sensitivity of EW precision tests to NP effects is illustrated by the constraints obtained on oblique parameters and on Wilson coefficients in the SMEFT framework. With the increasing accuracy of LHC measurements and future collider projects, precision tests of the SM will continue to play a crucial role in probing NP effects beyond the reach of direct searches. On one hand, precision tests might lead us to an indirect discovery of NP effects; on the other hand, should NP be directly observed at the next generation of high-energy colliders, precision tests will be essential to characterise its properties and unveil its nature. Combined theoretical and experimentail efforts aimed at improving the accuracy of global precision fits will therefore remain a central activity in particle physics for the years to come.

\begin{ack}[Acknowledgments]%
It is a pleasure to thank Jorge de Blas, Marco Ciuchini, Victor Miralles, Satoshi Mishima, Maurizio Pierini, and Mauro Valli for collaborating on several projects related to this review. The work of L.R. is supported in part by the U.S. Department of Energy under grant DE- SC0010102 and by
INFN through a Foreign Visiting Scientist Fellowship. This work was supported in part by the European Union - Next Generation
EU under italian MUR grant PRIN-2022-RXEZCJ.
\end{ack}

%%%%%%%%%%%%%%%%%%%%%%%%%%%%%%%%%%%%%%%%%%%%
%% Optional: A list of references to other relevant works/articles/websites which are not cited in the text but that would further enhance a readers understanding of this topic
%\seealso{article title article title}

%%%%%%%%%%%%%%%%%%%%%%%%%%%%%%%%%%%%%%%%%
%% Mandatory: Bibliography using bibtex 
\bibliographystyle{Numbered-Style} %% for Numbered Reference Style
\bibliography{hepbiblio}

\end{document}